\documentclass[twocolumn,aps,prd,superscriptaddress,nofootinbib]{revtex4-1}
\usepackage{graphicx}
\usepackage{amsmath}
\usepackage{bm}
\usepackage{slashed}
\usepackage{epsfig}
\usepackage{amsfonts}
\usepackage{epstopdf}
\usepackage{hyperref}
\usepackage[subfigure]{graphfig}
\usepackage{braket}
\usepackage{tikz}

\newcommand{\ben}{\begin{eqnarray}}
\newcommand{\een}{\end{eqnarray}}

\newcommand{\bef}{\begin{figure}[!htp]}
\newcommand{\eef}{\end{figure}}

\newcommand{\nn}{\nonumber}
\newcommand{\Tr}{{\rm Tr}}
\newcommand*\circled[1]{\tikz[baseline=(char.base)]{
            \node[shape=circle,draw,inner sep=2pt] (char) {#1};}}

\def\be{\begin{equation}}
\def\ee{\end{equation}}
\newcommand{\bea}{\begin{eqnarray}}
\newcommand{\eea}{\end{eqnarray}}

\def\ba{\begin{linenomath*}\begin{equation}}
\def\ea{\end{equation}\end{linenomath*}}

\usepackage{soul}

\hypersetup{colorlinks, linkcolor = [rgb]{0,0.0,1.0}, citecolor = [rgb]{0,0.0,1.0}, urlcolor = [rgb]{0,0.0,1.0}}

\begin{document}

\null\hfill\begin{tabular}[t]{l@{}}
  {JLAB-THY-19-2847} 
\end{tabular}
\preprint{JLAB-THY-19-2847}

\title{{Pion Valence Quark Distribution from  Matrix Element Calculated in Lattice QCD} }

\newcommand*{\JLAB}{Thomas Jefferson National Accelerator Facility, Newport News, VA 23606, USA}\affiliation{\JLAB}
\newcommand*{\WM}{Physics Department, College of William and Mary, Williamsburg, Virginia 23187, USA}\affiliation{\WM}

\author{Raza Sabbir Sufian}\affiliation{\JLAB}
\author{Joseph Karpie}\affiliation{\WM}
\author{Colin Egerer}\affiliation{\WM}
\author{Kostas Orginos}\affiliation{\JLAB}\affiliation{\WM}
\author{Jian-Wei Qiu}\affiliation{\JLAB}
\author{David G. Richards}\affiliation{\JLAB}


\begin{abstract}
We present the first exploratory lattice QCD calculation of the pion valence quark distribution extracted from spatially separated current-current correlations in coordinate space. We show that an antisymmetric combination of vector and axial-vector currents provides direct information on the pion valence quark distribution. Using the collinear factorization approach, we calculate the perturbative tree-level kernel for this current combination and extract the pion valence distribution. The main goal of this article is to demonstrate the efficacy of this general lattice QCD approach in the reliable extraction of  parton distributions. With controllable power corrections and a good understanding of the lattice systematics, this method has the potential to serve as a complementary to the many efforts to extract parton distributions in global analyses from experimentally measured cross sections. We perform our calculation on an ensemble
of 2+1 flavor QCD using the isotropic-clover fermion action, with lattice dimensions $32^3\times 96$ at a lattice spacing \mbox{$a=0.127$ fm} and the quark mass equivalent to a pion mass $m_\pi \simeq 416$ MeV. 
\end{abstract}

\maketitle
\allowdisplaybreaks


\section{Introduction}	
\label{sec:intro}

In the hard scattering processes involving hadrons, such as in the deep inelastic scattering (DIS) of leptons on hadrons, the experimentally measured cross sections are a combination of short- and long-distance physics. The inclusive DIS cross section can be factorized into a short-distance partonic hard part 
which is calculable order by order in perturbation theory and a 
long-distance hadronic part which can be represented by universal and nonperturbative distribution functions, called the parton distribution functions (PDFs), plus corrections suppressed by inverse power of large momentum transfer of the scattering. It is the QCD factorization theorem~\cite{Collins:1989gx} which enables us to connect the dynamics of quarks and
gluons to the physically measured hard scattering cross sections of identified hadrons. The collinear (CO) divergences of the partonic scattering
are absorbed into the nonperturbative PDFs, leaving an infrared-safe and perturbatively computable hard contribution. According to Feynman's parton model~\cite{Feynman}, the unpolarized PDFs give the probability to find partons [{\em i.e.} quark ($q$), antiquark ($\bar{q}$), gluon ($g$)] in a hadron as a function of the fraction $x$ of the hadron's longitudinal momentum carried by the parton, probed at a factorization scale $\mu$. For example, if a parton of type $i$ carries a fraction $x$ of hadron's momentum, then the probability to find the parton is given by $f_{i}(x,\mu^2)dx$. An accurate and precise knowledge of parton distribution functions is required for the cross section predictions of both Standard Model and Beyond Standard Model processes at existing and future particle colliders, such as the LHC and Electron-Ion Collider (EIC). The precision of numerous experimentally measured observables, such as the $W$-boson mass, weak-mixing angle and Higgs cross section, is driven by a detailed knowledge and precision of PDFs.  \\

Since a precise knowledge of PDFs is required for the analysis and interpretation of scattering experiments, as discussed above, considerable effort has been made to determine PDFs and their uncertainties by global fitting collaborations such as MMHT~\cite{Harland-Lang:2014zoa}, CT~\cite{Dulat:2015mca}, NNPDF~\cite{Ball:2017nwa},  HERAPDF~\cite{Alekhin:2017kpj}, and JAM~\cite{Ethier:2017zbq}. PDFs, the catalyst of many observables in hadronic scattering, are becoming better determined as experimental data sets increase and the global analysis community implements more sophisticated schemes to quantify systematic uncertainties. 

The valence quark distribution of the pion is of particular theoretical interest, as the pion is the lightest QCD bound state and the Goldstone mode associated with dynamical chiral symmetry breaking. The pion PDF has been measured through pionic Drell-Yan experiments at CERN~\cite{Badier:1983mj,Betev:1985pf} and Fermilab~\cite{Conway:1989fs}. Several analyses in Refs.~\cite{Owens:1984zj,Aurenche:1989sx,Sutton:1991ay,Gluck:1991ey,Wijesooriya:2005ir,Aicher:2010cb,Barry:2018ort} of these experimental data have been performed to determine the pion valence distribution. Among these analyses,  it has been emphasized in Ref.~\cite{Aicher:2010cb} that the next-to-leading-logarithmic threshold resummation effects in the calculation of the Drell-Yan cross section are important and give a softer valence distribution which falls of as $(1-x)^2$ near $x\to 1$, consistent with the prediction based on the framework of perturbative QCD  in Refs.~\cite{Farrar:1979aw,Berger:1979du,Brodsky:1994kg}. There are also different model predictions for the large-$x$ behavior of the pion valence distribution, some of which predict a harder fall-off as  $(1-x)$~\cite{Shigetani:1993dx,Davidson:1994uv,Melnitchouk:2002gh,deTeramond:2018ecg} or $(1-x)^2$,  such as in Dyson-Schwinger type models~\cite{Hecht:2000xa,Chen:2016sno}. Therefore,  an {\it ab initio} knowledge of the correct large-$x$ behavior of the pion valence PDF can serve as a discriminator of different model calculations. In this paper, we will present a calculation of the pion valence PDF
using ``lattice cross sections"  proposed in Refs.~\cite{Ma:2014jla,Ma:2017pxb}. In this approach, one factorizes a hadronic matrix element, such as a two current correlator, into the PDFs and  short-distance matching coefficients, from which PDFs could be extracted from lattice calculated hadronic matrix element with the perturbatively calculated matching coefficients.

Lattice QCD is a Monte Carlo method for numerically evaluating QCD in a finite, discretized Euclidean spacetime. To date, Lattice QCD has emerged as the most rigorous and systematic tool for studying QCD nonperturbatively. As introduced by Feynman~\cite{Feynman:1969ej}, PDFs are defined through light cone matrix elements of certain bilocal operators. These light cone matrix elements cannot be directly calculated on the Euclidean lattice because the light cone collapses to a point in Euclidean spacetime. Recently several methods have been introduced to go beyond the calculations of the first few moments of PDFs on the lattice, such as the path-integral formulation of the deep-inelastic scattering hadronic tensor~\cite{Liu:1993cv,Liu:1999ak}, the inversion method~\cite{Horsley:2012pz}, quasi-PDFs~\cite{Ji:2013dva}, and pseudo-PDFs~\cite{Radyushkin:2017cyf} to obtain the $x$-dependent hadron structure functions. A coordinate-space method for the calculation of light-cone distribution amplitudes has also been employed~\cite{Braun:2007wv}. Significant achievements in the lattice QCD implementations of these approaches have been made in recent years~\cite{Bali:2018spj,Chambers:2017dov,Alexandrou:2018pbm,Lin:2018qky,Orginos:2017kos}. References to many other lattice QCD calculations, the current status and challenges for a meaningful comparison of these lattice calculations with the global fits of PDFs can be found in Refs.~\cite{Lin:2017snn,Monahan:2018euv,Cichy:2018mum}. The readers are also referred to several lattice QCD calculations of lower moments of pion PDF in Refs.~\cite{Best:1997qp,Detmold:2003tm,Guagnelli:2004ga,Capitani:2005jp,Bali:2013gya,Abdel-Rehim:2015owa,Oehm:2018jvm}. Recently, the  relation  of  moments  of  Ioffe time parton  distribution  functions  to matrix elements of nonlocal operators computed in lattice QCD has been studied in~\cite{Karpie:2018zaz}. 

The remainder of this article is organized as follows.\\
In Secs.~\ref{GLCSs} and~\ref{Essence} we discuss what are the good lattice cross sections, and the essence of calculation of these lattice cross sections in coordinate space.  In Sec.~\ref{LOkernel}, we present the derivation of the tree-level perturbative kernel for an antisymmetric vector and axial-vector current combination, and show how one can factorize the associated hadronic matrix element to extract pion valence distribution in a lattice QCD calculation. We present the numerical methods and results in Secs.~\ref{methods} and~\ref{results}. We compare the pion valence distribution extracted in this calculation with other calculations and different fits of the experimental data. Finally, we summarize our results and outline the future directions of this method to obtain pion valence quark distribution with controlled systematics.


\section{``Good" Lattice Cross Sections} 
\label{GLCSs}

In similarity with the extraction of PDFs in a global fit through the factorization of different experimental cross sections, a method was proposed~\cite{Ma:2014jla} to extract PDFs from lattice QCD calculations of hadronic matrix elements, called lattice cross sections (LCSs), which in the framework of QCD can be factorized into a perturbatively calculable hard part and the nonperturbative PDFs with a small and controllable power correction. Analogous to the cross sections measured in an experiment, these hadronic matrix elements computed on the lattice are constructed to be time independent, defined by equal-time operators and have a well-defined continuum limit; hence the name lattice ``cross sections." It has been shown in Ref.~\cite{Briceno:2017cpo} that as long as such operators have no temporal extent, a matrix element calculated in Euclidean space will equal its counterpart in Minkowski space. 

To be more specific, such hadronic matrix elements are good lattice cross sections if they have the following properties:
\begin{itemize}
\item are a Lorentz covariant single-hadron matrix element computable on a Euclidean lattice,
\item have a well-defined continuum limit when the lattice spacing $a\to 0$,
\item are factorizable to PDFs convoluted with infrared (IR)-safe hard coefficients, plus a small and controllable power correction.
\end{itemize}

The single-hadron matrix elements of renormalized nonlocal operators ${\cal O}_n({\xi})$ can be written as,  suppressing renormalization scale dependence, 
\begin{align}\label{eq:lcs}
{\sigma}_{n}(\xi,p)=\langle p| {T}\{{\cal O}_n({\xi})\}|p\rangle,
\end{align}
where the subscript $n$ is a label for different operators, $T$ stands for time-ordering, $p$ the hadron momentum, and $\xi$  ($\xi^2\neq0$ ) is the largest separation of all fields in the operator ${\cal O}_n$. One choice which allows for factorization is a pair of parton field operators linked by a Wilson line operator and has been used for the calculation of pseudo-PDFs~\cite{Radyushkin:2017cyf} and quasi-PDFs~\cite{Ji:2013dva}. A broader class of operators with factorizable matrix elements is pairs of spacelike separated currents. 

The operator can be chosen to be a Lorentz scalar, such as
\begin{align}\label{eq:2scalarcurrents}
{\cal O}_{j_1j_2}(\xi)\equiv
&\,
\xi^{d_{j_1}+d_{j_2}-2}\, Z_{j_1}\, Z_{j_2} j_1(\xi/2)\, j_2(-\xi/2)\, ,
\end{align}
where $j_1$ and $j_2$ are currents with no Lorentz indices such as $\bar\psi\psi$ or $\bar\psi \slashed\xi\psi$, $d_j$ and $Z_j$ are the dimension and renormalization constant of the current $j$, respectively, and the overall dimensional factor is introduced so that the matrix elements in Eq.~\eqref{eq:lcs} are dimensionless with our normalization, $\langle p|p'\rangle = (2E_p)(2\pi)^3\delta^3(p-p')$. In this case the LCS can be written in terms of only Lorentz invariants 
\begin{align} \label{eq:scalarlcs}
{\sigma}_{j_1 j_2}(\omega, \xi^2)=\langle p| {T}\{{\cal O}_{j_1 j_2}({\xi})\}|p\rangle,
\end{align}
where the Lorentz scalar $\omega\equiv p\cdot\xi$ is the Ioffe time~\cite{Ioffe:1969kf}, and $p^2$ dependence is suppressed. This type of LCS can be directly factorized into the PDF when the separation $|\xi|$ is small~\cite{Ma:2014jla}.   

Other choices of operators can have a more complicated Lorentz structure, such as the vector-vector matrix element of the operator
\begin{align}\label{eq:2vectorcurrents}
{\cal O}^{\mu\nu}_{VV}(\xi)\equiv
&\,
\xi^{4}\, j_V^\mu(\xi/2)\, j_V^\nu(-\xi/2)\, ,
\end{align}
where $j_V^\mu$ is the vector current which requires no renormalization constants. This type of LCS will need to be decomposed into the Lorentz structures allowed by symmetry. The functions of Lorentz invariants which accompany these Lorentz structures are the objects which will be factorized into PDFs. For the case of the operator ${\cal O}^{\mu\nu}_{VV}$, its matrix element of an unpolarized hadron state is symmetric in $\{\mu,\nu\}$ and can be decomposed as,
\begin{align*}
\sigma^{\mu\nu}_{VV}(\xi,p) = p^\mu p^\nu T_1(\omega,\xi^2)  + \frac 12 ( p^\mu \xi^\nu + \xi^\mu p^\nu) T_2(\omega,\xi^2) \\+  g^{\mu\nu} T_3(\omega,\xi^2)+  \xi^\mu \xi^\nu T_4(\omega,\xi^2)
\end{align*}
where $T_i$ are Lorentz invariant functions. In the following section, we discuss the vector and axial-vector current combination which will be used to extract the pion valence quark distribution. \\

It is worth noting that the LCSs have the following analogs to hard scattering experiments:
\begin{itemize}
\item 
the label ``$n$'' in Eq.~(\ref{eq:lcs}) 
is related to the dynamical features of LCSs and mimics different processes in experiments.
\item $p$ and $\xi$ are analogous to 
observed scales defining 
the collision kinematics; $p$ relating to the collision energy $\sqrt{S}$ and $\xi^2$ relating to the hard probe $\frac{1}{Q^2}$.
\end{itemize}


\section{The Essence of Calculation in Coordinate Space } 
\label{Essence}

It is crucial to mention that a large $p$ alone does not guarantee the applicability of  the operator product expansion of the matrix element and contributions from large $\xi$ can invalidate the perturbative factorization, whether for the case of quark-antiquark fields linked by a Wilson line, or for the case of spatially-separated currents. The validity of perturbative factorization requires that the separation's scale, $\xi^{2}$, be much smaller than the inverse square of typical hadronic scale, $\Lambda_{\rm QCD}$, namely $\xi^2 \Lambda^2_{\rm QCD} \ll 1$. 

We now show why the coordinate space approach provides distinct theoretical advantages. One can write the Fourier transform of $\sigma_n(\omega, \xi^2 )$:
\bea \label{FT}
\widetilde{\sigma}_{n}(\widetilde{\omega},q^2)\equiv
\int \frac{d^4\xi}{\xi^4} e^{iq\cdot \xi} {\sigma}_{n}(\omega,\xi^2 ),
\eea
where corresponding ${\cal O}_n$ can be any operator defined in Eq.~\eqref{eq:lcs}, and $\widetilde{\omega}\equiv \frac{2p\cdot q}{-q^2}=\frac{1}{x_{\rm B}}$ with $x_{\rm B}$ the Bjorken variable for the lepton-hadron DIS.
However, though $q$ is related to $\xi$ through the Fourier transform above, it is not a one-to-one relation. $\tilde\sigma$ with small or large values of $q$ will receive contribution from $\sigma$ with all values of $\xi$, small and large. In particular,  it can involve contributions from values of $\xi\gg \frac{1}{\Lambda_{\rm QCD}}$, thereby violating factorization. This is why, in the LCSs approach, $\xi$ is a very well-defined quantity, analogous to a hard probe of hadron structure in a DIS experiment. It has also been demonstrated in Ref.~\cite{Ma:2017pxb} that, the non-analytic cut of $\widetilde{\sigma}_{n}$ comes from the integration region of large $\xi$. That is, even if we demand $|q^2|\gg \Lambda_{\rm QCD}^2$, $\widetilde{\sigma}_{n}$ in momentum space can always receive contribution from large $\xi$ region so long as $\widetilde{\omega}^2>1$.  On the other hand, in coordinate space, if we fix $\xi$ to be short-distance, we do not have contribution from the large $\xi$ region and thus ${\sigma}_{n}$ has a good analytic behavior.


\section{Factorization} 
\label{LOkernel}

The lattice calculable hadronic matrix elements of Eq.~\eqref{eq:lcs} are shown in Ref.~\cite{Ma:2017pxb} to be factorizable into PDFs with perturbatively calculable coefficients by applying the operator product expansion (OPE) to the nonlocal operator $\mathcal{O}_n\left(\xi\right)$, with small but nonvanishing $\xi^2$
\bea
\sigma_n^h(\omega,\xi^2,p^2) &=&\sum_a\int^1_{-1}\frac{dx}{x}f^h_a (x,\mu^2)\nn \\  
&&\times K_n^a (x\omega,\xi^2,x^2p^2,\mu^2) + \mathcal{O}\left(\xi^2\Lambda_{\text{QCD}}^2\right),\nn \\
\label{eq:factor}
\eea
where $\sigma_n^h$ is $\mathcal{O}_n\left(\xi\right)$ measured in a hadron $h$, $K_n^a$ are the parton flavor $a\in\lbrace q,\overline{q},g\rbrace$ contributions to the perturbative hard coefficients with corresponding PDF $f^h_a$, and factorization scale $\mu^2$. The short-distance coefficient functions
$K_n^a\left(\omega,\xi^2;p^2,\mu^2\right)$
are determined by applying the factorized formula in Eq.~\eqref{eq:factor} to an asymptotic parton of momentum $p$ with $p^2=0$ and flavor $a=q, \overline{q}$, or $g$, and expanding each side as a power series in the strong coupling constant $\alpha_s$. Although the perturbative coefficient functions are process-dependent, they apply equally to different external hadron states. At leading-order $\mathcal{O}\left(\alpha_s\right)$, the matching coefficient only receives quark contributions and the factorized relation in Eq.~\eqref{eq:factor} becomes
\bea
\sigma_n^{q\left(0\right)}\left(\omega,\xi^2\right)&=&
\sum_{a=q,\overline{q},g}\int_0^1\frac{dx}{x}f_a^{q\left(0\right)}\left(x,\mu^2\right)\nn \\
&&\times K_n^{a\left(0\right)}\left(x \omega,\xi^2;\mu^2\right)\!+\!\mathcal{O}\left(\xi^2\Lambda^2_{\rm QCD}\right),
\label{eq:LO}
\eea
where $p^2=0$ is suppressed and
$f_a^{q\left(0\right)}\left(x,\mu^2\right)=\delta\left(1-x\right)\delta^{qa}$
is the quark distribution of an asymptotic quark at zeroth order in $\alpha_s$ and does not have the factorization scale $\mu^2$-dependence.
Upon substitution of $f_a^{q\left(0\right)}$ into
Eq.~\eqref{eq:LO}, it can be shown
\be
\sigma_n^{q\left(0\right)}\left(\omega,\xi^2\right)=K_n^{q\left(0\right)}\left(\omega,\xi^2\right).
\label{eq:kernel-matelem-equality}
\ee
where the active quark momentum $p^2=0$ is suppressed.
Thus determination of the leading-order coefficient functions $K_n^{q\left(0\right)}$ follows directly from the expression of the matrix elements $\sigma_n^{q\left(0\right)}\left(\omega,\xi^2\right)$ in the coordinate space.

Specializing to the case of the pion, consider a generic tensor operator
 \be
  \mathcal{O}_{ij}^{\mu\nu}\left(\xi\right)=\xi^4\mathcal{J}^\mu_i\left(\xi/2\right)\mathcal{J}^\nu_j\left(-\xi/2\right)\, ,      
  \label{eq:operator}                                                                                  
  \ee
  and have it evaluated in the pion state $\ket{\pi\left(p\right)}$, where $\mathcal{J}_k$ is a local quark bilinear and $\xi^4$ is included to maintain an overall dimensionless matrix element. By examining the path-integral definition of an arbitrary operator defined at a single Euclidean time removes complications in analytically continuing our results back to Minkowski space. In this case, the general time-ordering of $\mathcal{O}_{ij}^{\mu\nu}\left(\xi\right)$ is instead expressed as a sum of diagrams with momenta flowing in/out of the fixed current locations. We define the matrix element of $\mathcal{O}_{ij}^{\mu\nu}\left(\xi\right)$ in the pion as
\bea
\sigma^{\mu\nu}_{ij}\left(\xi,p\right) &&=\bra{\pi\left(p\right)}\mathcal{O}^{\mu\nu}_{ij}\left(\xi\right)\ket{\pi\left(p\right)}\nn \\
&&=\xi^4\bra{\pi\left(p\right)}\mathcal{J}^\mu_i\left(\xi/2\right)\mathcal{J}^\nu_j\left(-\xi/2\right)\ket{\pi\left(p\right)}
\label{eq:sigma-q}
\eea
Projecting onto an asymptotic quark state, we are left with two distinct diagrams at leading order (LO):
%

\begin{figure}[h]
  \centering
  \subfigure[]{\includegraphics[width=0.2\textwidth]{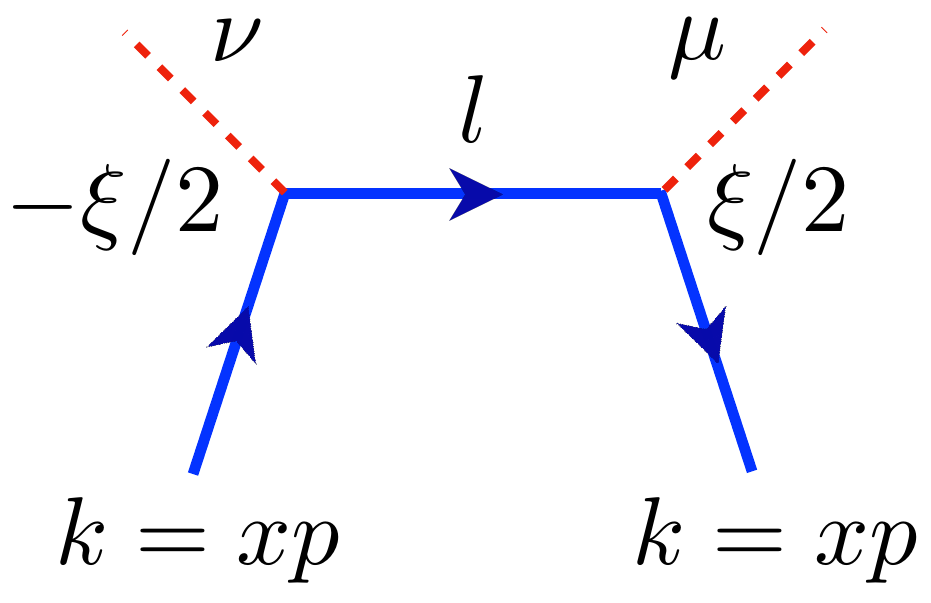}\label{a}}
  \subfigure[]{\includegraphics[width=0.2\textwidth]{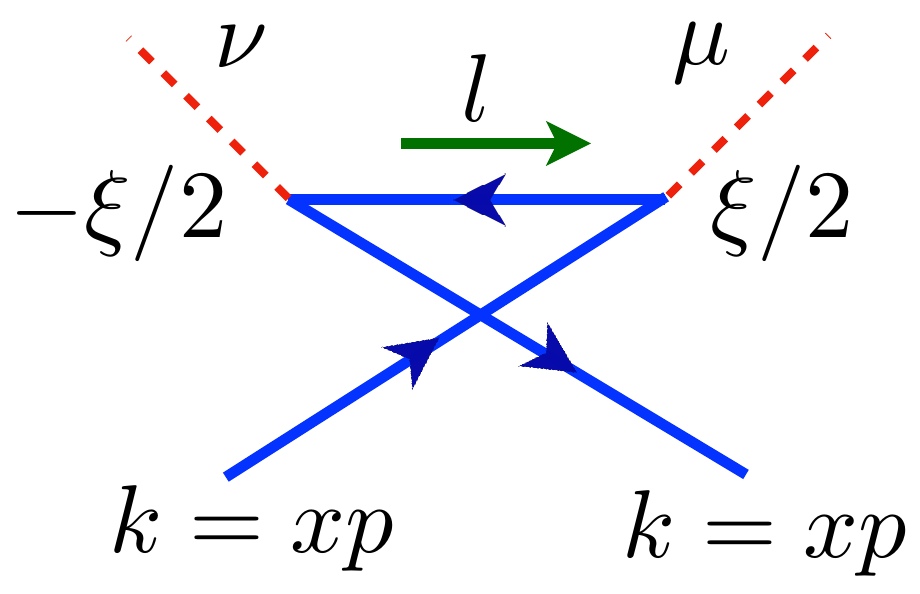}\label{b}}
    \caption{\label{fig:LOdiags}
    The lowest order Feynman diagrams contributing to the $\sigma^{\mu\nu}_{ij}$ in Eq.(\ref{eq:sigma-q}) on an asymptotic on-shell quark state of momentum $k$.}
\end{figure}

Depending on the current-current combinations considered, the resulting Lorentz decomposition of $\sigma^{\mu\nu}_{ij}\left(\xi,p\right)$ will introduce numerous scalar form factors consistent with parity and time-reversal invariance. It is these form factors that will provide information on a wide array of distribution functions, when factorized according to Eq.~\eqref{eq:factor}. A general expression from which the LO perturbative kernels can be obtained follows from application of perturbative formulae to the diagrams above. Averaging over quark spin, the ordering depicted in Fig.~(\ref{a}) yields
\bea
\mathcal{M}_{ij}^{(a)}   &&=\frac{\xi^4}{2}\sum_s\bra{0}\overline{u}^s\left(k\right)e^{ik\cdot\xi/2} \Gamma_{i}^\mu \psi\left(\xi/2\right)\nn \\
&&\times~\overline{\psi}\left(-\xi/2\right) \Gamma_{j}^\nu e^{ik\cdot\xi/2}u^s\left(k\right)\ket{0}\nn \\
&&=\frac{\xi^4}{2}\! \sum_se^{ik\cdot\xi}\overline{u}^s\left(k\right)\Gamma_i^\mu\! \bra{0}\psi\left(\xi/2\right)\overline{\psi}\left(-\xi/2\right)\ket{0}\! \Gamma_j^\nu u^s\left(k\right)\nn \\
&&=\frac{\xi^4}{2}e^{ik\cdot\xi}~\Tr{\left[\left(\gamma\cdot k\right)\Gamma_i^\mu\int\frac{d^4l}{\left(2\pi\right)^4}\frac{i\gamma\cdot l}{l^2+i\epsilon}e^{-il\cdot\xi}~\Gamma_j^\nu\right]}
\label{eq:wick1}
\eea
where an inverse Fourier transform has been used to express the quark propagator from $-\xi/2\rightarrow\xi/2$ in coordinate space. The second ordering, shown in Fig.~(\ref{b}), similarly yields
\be
\mathcal{M}_{ji}^{(b)}=\frac{\xi^4}{2}e^{-ik\cdot\xi}~\Tr{\left[\left(\gamma\cdot k\right)\Gamma_j^\nu\int\frac{d^4l}{\left(2\pi\right)^4}\frac{-i\gamma\cdot l}{l^2+i\epsilon}e^{-il\cdot\xi}~\Gamma_i^\mu\right]}
\label{eq:wick2}
\ee

Combining Eqs.~\eqref{eq:wick1} and~\eqref{eq:wick2} and writing the quark momentum as $k_\mu=xp_\mu$, we obtain a general relation in the LO denoted by the superscript $(0)$ as 
\bea
\sigma^{\mu\nu\left(0\right)}_{ij}\left(p\cdot\xi,p;x,\xi\right)  &=&\frac{i}{4\pi^2}xp_\alpha\xi_\beta\left\{e^{ixp\cdot\xi}~\Tr{\left[\gamma^\alpha\Gamma_i^\mu\gamma^\beta\Gamma_j^\nu\right]}\right.\nn \\
  &&\left. -e^{-ixp\cdot\xi}~\Tr{\left[\gamma^\alpha \Gamma_j^\nu \gamma^\beta \Gamma_i^\mu \right]}\right\}
\label{eq:master-lo-kernel}
\eea
from which the kernels $K_n^{q\left(0\right)}\left(\omega,\xi^2;x\right)$
with $\omega=p\cdot\xi$
can be isolated for currents $\lbrace i,j\rbrace$.

Given invariance of the strong interaction under parity ($\mathcal{P}$) and time-reversal ($\mathcal{T}$) transformations, the pion matrix element $\sigma_{ij}^{\mu\nu}\left(\xi,p\right)$ has the following property,
\be
\sigma^{\mu\nu}_{ij}\left(\xi,p\right)=\bra{\pi\left(p\right)}\left(\mathcal{P}\mathcal{T}\right)\left(\mathcal{O}^{\mu\nu}_{ij}\left(\xi\right)\right)^\dagger\left(\mathcal{P}\mathcal{T}\right)^{-1}\ket{\pi\left(p\right)}.
\ee

In this work, we consider the case of a vector $\mathcal{J}_V^\mu=\overline{\psi}\gamma^\mu\psi$ and axial-vector $\mathcal{J}_A^\nu=\overline{\psi}\gamma^\nu\gamma^5\psi$ current combination, whose transformation properties are
\bea
&&\left(\mathcal{P}\mathcal{T}\right)\mathcal{J}^\mu_A\left(\xi\right)\left(\mathcal{P}\mathcal{T}\right)^{-1}=-\mathcal{J}_A^\mu\left(-\xi\right)\nn \\
&&\left(\mathcal{P}\mathcal{T}\right)\mathcal{J}^\mu_V\left(\xi\right)\left(\mathcal{P}\mathcal{T}\right)^{-1}=\mathcal{J}_V^\mu\left(-\xi\right)\nn
\eea
With these transformation properties, we find that the following combination of these two currents, $\sigma_{VA}^{\mu\nu}\left(\xi,p\right)+\sigma^{\mu\nu}_{AV}\left(\xi,p\right) \equiv\bra{\pi\left(p\right)}\left[\mathcal{O}^{\mu\nu}_{VA}\left(\xi\right)+\mathcal{O}_{AV}^{\mu\nu}\left(\xi\right)\right]\ket{\pi\left(p\right)}$, is antisymmetric in Lorentz indices, $\{\mu,\nu\}$, and can be expressed in terms of two dimensionless scalar form factors as
\bea
&&\frac{1}{2}\left[\sigma_{VA}^{\mu\nu}\left(\xi,p\right)+\sigma^{\mu\nu}_{AV}\left(\xi,p\right)\right]\nn \\
&&\quad\equiv\epsilon^{\mu\nu\alpha\beta}\xi_\alpha p_\beta T_1\left(\omega,\xi^2\right)+\left(p^\mu\xi^\nu-\xi^\mu p^\nu\right)T_2\left(\omega,\xi^2\right)\nn \\
\label{eq:symcomb}
\eea
where $T_i\left(\omega,\xi^2\right)$ are the dimensionless functions of the Lorentz invariants $\lbrace\omega,\xi^2\rbrace$.

The dimensionless functions are isolated by taking appropriate tensor contractions of the antisymmetric matrix element in Eq.~(\ref{eq:symcomb}),
\bea
T_1\left(\omega,\xi^2\right)\!&=&\!\frac{1}{2\left(\omega^2-p^2\xi^2\right)}  
\label{eq:T1} \\
&\times & \left(\epsilon_{\mu\nu\alpha\beta} \xi^\alpha p^\beta \right)\! 
\frac{1}{2}\left[\sigma_{VA}^{\mu\nu}\left(\xi,p\right)+\sigma_{AV}^{\mu\nu}\left(\xi,p\right)\right]\, , 
\nn \\
T_2\left(\omega, \xi^2\right)\!&=&\!\frac{1}{2\left(\omega^2-p^2\xi^2\right)} 
\label{eq:T2}\\
&\times & \left(\xi_\mu p_\nu-p_\mu\xi_\nu\right)\!\frac{1}{2}\!
\left[\sigma_{VA}^{\mu\nu}\!\left(\xi,p\right)\!+\!\sigma_{AV}^{\mu\nu}\!\left( \xi,p\right)\right] \, . \nn
\eea

A judicious choice of $\xi$, $p$, and Lorentz indices $\{\mu,\nu\}$, exposes the structure functions $T_1\left(\omega,\xi^2\right)$ and $T_2\left(\omega,\xi^2\right)$ without recourse to a full tensor contraction as in Eqs.~\eqref{eq:T1} and~\eqref{eq:T2}. To isolate the structure functions we stipulate $p=(p^0, 0, 0, p^3)$ and $\xi = (0,0,0,\xi^3)$. $T_1$ is then isolated by choosing $\mu=1$ and $\nu=2$:
\bea
T_1\left(\omega,\xi^2\right)\!&=&\!\frac{1}{p^0\xi^3} \frac{1}{2}\left[\sigma^{12}_{VA}\left(\xi,p\right)+\sigma^{12}_{AV}\left(\xi,p\right)\right] \,. 
\eea
While $T_2$ is isolated by choosing $\mu=0$ and $\nu=3$:
\bea
T_2\left(\omega, \xi^2\right)\!&=&\!\frac{1}{p^0 \xi^3}\!\frac{1}{2}\!\left[\sigma^{03}_{VA}\!\left(\xi,p\right)\!+\!\sigma^{03}_{AV}\!\left( \xi,p\right)\right] \, .
\label{eq:FFs}
\eea
From Eq.~(\ref{eq:factor}), the $T_i\left(\omega,\xi^2\right)$ can be thus factorized as
\bea
T_i\left(\omega,\xi^2\right)&=&\sum_{a=q,\overline{q},g}\int^1_0\frac{dx}{x} f_a \left(x,\mu^2\right) C^a_i \left(x\omega,\xi^2,\mu^2 \right)\nn \\
&&+\mathcal{O}\big(\xi^2\Lambda^2_{\rm QCD}\big). 
\label{eq:Ti-kernels}
\eea
with the perturbatively calculable matching coefficients $C^a_i \left(x\omega,\xi^2,\mu^2 \right)$ for parton flavor $a$.

    Similar to the derivation of Eq.~(\ref{eq:master-lo-kernel}), the LO contribution to the antisymmetric pion matrix element in Eq.~(\ref{eq:symcomb}) is given by
\bea
\sigma^{\mu\nu\left(0\right)}_{VA}  &&\left(p\cdot\xi,p;x,\xi\right)=\frac{i}{4\pi^2}xp_\alpha\xi_\beta\left[\Tr{\big(\gamma^\alpha\gamma^\mu\gamma^\beta\gamma^\nu\gamma^5\big)}e^{ixp\cdot\xi}\right.\nn \\
  &&\qquad\qquad\qquad~\left.-\Tr{\big(\gamma^\alpha\gamma^\nu\gamma^5\gamma^\beta\gamma^\mu\big)}e^{-ixp\cdot\xi}\right]\nn \\
&&\qquad\qquad=-\frac{i}{\pi^2}xp_\alpha\xi_\beta\left(i\epsilon^{\alpha\mu\beta\nu}e^{ix\omega}-i\epsilon^{\alpha\nu\beta\mu}e^{-ix\omega}\right)\nn \\
&&\qquad\qquad=\frac{1}{\pi^2}x\epsilon^{\mu\nu\alpha\beta} \xi_\alpha p_\beta\left(e^{ix\omega}+e^{-ix\omega}\right)
\label{eq:VA-kernel}
\eea
with $\Tr{\big(\gamma^\mu\gamma^\nu\gamma^\rho\gamma^\sigma\gamma^5\big)}=-4i\epsilon^{\mu\nu\rho\sigma}$ dictated by the convention $\epsilon^{0123}=1$. Substituting Eq.~(\ref{eq:VA-kernel}) into Eqs.~(\ref{eq:T1}) and (\ref{eq:T2}), we obtain the two scalar form factors of a quark state of momentum $k=xp$ at the LO, respectively,
  \bea
  T_1^{q(0)}(x\omega,\xi^2) &=&
  \frac{x}{\pi^2}\left(e^{ix\omega}+e^{-ix\omega}\right)\, ,
  \label{eq:T10} \\
  T_2^{q(0)}(x\omega,\xi^2) &=& 0 \, .
  \label{eq:T20} 
  \eea
  With $f_a^{q\left(0\right)}\left(x,\mu^2\right)=\delta\left(1-x\right)\delta^{qa}$, we obtain the lowest order perturbative coefficients in Eq.~(\ref{eq:Ti-kernels}) as,
  \bea
  C_1^{q(0)}(x\omega,\xi^2) &=& T_1^{q(0)}(x\omega,\xi^2) 
  = \frac{2x}{\pi^2}\, \cos(x\omega)\,
  \label{eq:C10} \\
  C_2^{q(0)}(x\omega,\xi^2) &=& 0 \, ,
  \label{eq:C20} 
  \eea
  respectively.

  We have the LO momentum-space scalar form factors by performing a Fourier transformation in $\omega$,
  \bea
  \tilde{T}_1  &&\left(\tilde{x},\xi^2\right)\equiv \int\frac{d\omega}{2\pi}e^{-i\tilde{x}\omega}\, T_1\left(\omega,\xi^2\right)\nn \\
  &&\quad\approx\int\frac{d\omega}{2\pi}e^{-i\tilde{x}\omega}\int_0^1\frac{dx}{x}q\left(x\right) C_1^{q(0)}(x\omega,\xi^2,\mu^2) \nn\\
  &&\quad\approx\int\frac{d\omega}{2\pi}e^{-i\tilde{x}\omega}\int_0^1\frac{dx}{x}q\left(x\right)
  \frac{x}{\pi^2}\left(e^{ix\omega}+e^{-ix\omega}\right)\nn \\
  &&\quad\approx\frac{1}{\pi^2}\big\lbrace q\left(\tilde{x}\right)+q\left(-\tilde{x}\right)\big\rbrace\nn \\
  &&\quad\approx\frac{1}{\pi^2}\big\lbrace q\left(\tilde{x}\right)-\overline{q}\left(\tilde{x}\right)\big\rbrace 
  = \frac{1}{\pi^2}\, q_{\rm v}\!\left(\tilde{x}\right), 
  \label{eq:T1v}
  \eea
  where $q(-\tilde{x})=-\overline{q}(\tilde{x})$ is used, $q_{\rm v}\!\left(\tilde{x}\right)\equiv [q\left(\tilde{x}\right)-\overline{q}\left(\tilde{x}\right)]$ is the valence quark distribution, and the $\xi^2$ or the factorization scale dependence is suppressed since we are working at the LO approximation.  Equation~(\ref{eq:T1v}) implies that $\tilde{T}_1(\tilde{x}, \xi^2)$ is proportional to the valence quark PDF with momentum fraction $\tilde{x}$, which is actually true to all orders due to the symmetry of the coefficient function $C_1^q(x\omega,\xi^2,\mu^2) = - C_1^q(-x\omega,\xi^2,\mu^2)$.

Therefore direct information on the pion's valence quark distribution $q_{\rm v}\left(\tilde{x},\mu^2\right)$ is accessible by evaluating the antisymmetric combination of vector and axial-vector (V-A) current-current correlators, up to an overall factor of $1/\pi^2$ and corrections in powers of $\alpha_s$ and/or $\xi^2\Lambda_{\rm QCD}^2$.

It has been shown in Ref.~\cite{Ma:2017pxb} that the validity of operator product expansion (OPE) guarantees that $T_1$ is an analytic function of $\omega$, as is its Taylor series around $\omega=0$. By keeping $\xi$ to be short distance and increasing $\omega$ by increasing $p$, there exists no way for new divergences to appear in $T_1$. Therefore, $T_1$ remains an analytic function of $\omega$ unless $\omega=\infty$ and the factorization holds for any values of $\omega$ and $\xi^2$ as long as $\xi$ is short distance, similar to the scenario of the factorization of experimental cross sections. 


%
\section{Numerical Methods} 
\label{methods}

This calculation is performed on a lattice gauge ensemble of 490 configurations generated by the JLab/W\&M Collaboration~\cite{lattices}. This ensemble employs 2+1 flavors of clover Wilson fermions and a tree-level tadpole improved Symanzik gauge action. The strange quark mass was set by requiring the ratio $\left(2M^2_{K^+}-M^2_{\pi^+}\right)/M_{\Omega^-}$ to assume its physical value. The configurations were generated using a rational Hybrid Monte Carlo update algorithm~\cite{rhmc}. The fermion action includes a single iteration of stout smearing with weight $\rho = 0.125$. This smearing makes the employed tadpole corrected tree-level clover coefficient, $c_{sw}$, very close to the nonperturbative value determined,  {\it a posteriori}, by the Schr\"odinger functional method.

The extraction of hadron-to-hadron matrix elements in lattice QCD requires the calculation of correlation functions. The 2-point function is a vacuum expectation value of two interpolating fields separated in Euclidean time $T$:
\bea
C_{\rm 2pt}(p,T) = \langle \Pi_p(T) \overline{\Pi}_p(0)\rangle,
\eea
where the interpolating field $\Pi_p$ is an operator with quantum numbers of a pion with momentum $p$. A spectral decomposition of the 2-point function is given by the following tower of exponentials 
\bea
C_{\rm 2pt}(p,T) = \sum_n \frac{|Z_n|^2}{2E_n(p)} e^{-E_n(p) T} ,
\eea 
where the sum is over all energy eigenstates $n$ with quantum numbers of the pion, $Z_n = \langle 0 | \Pi_p|n\rangle $ is the overlap factor between the operator and the $n{\rm th}$ excited state and $E_n(p)$ is the energy of that state with momentum $p$. In the large Euclidean time limit, this correlation function will be dominated by the ground state. 

A good choice of interpolating field will have a large overlap factor with the ground state while simultaneously having poor overlap with excited states. For low-momenta or states at rest, spatial smearing is a well-established method to reduce the overlap of pointlike interpolators onto high energy eigenstates. We employ in this work the Jacobi-smearing procedure~\cite{Allton:1993wc}, in which pointlike quark fields are smeared according to
\be
\hat{q}\left(\vec{x},t\right)=\left(1+\frac{\sigma\nabla^2}{n_\sigma}\right)^{n_\sigma}q\left(\vec{x},t\right)
\ee
where $\nabla^2$ is the three dimensional gauge-covariant discretization of the Laplacian, $\sigma$ the smearing ``width'' and $n_\sigma$ the number of applications of the smearing kernel onto the pointlike quark fields.
For highly-boosted states, however, the overlap of even spatially-smeared interpolators can become suboptimal. To ameliorate the effects of excited-states and improve the overlap of our interpolators onto boosted pions, we implement a combination of the Jacobi and momentum-smearing~\cite{Bali:2016lva} techniques. In practice we apply appropriately constructed phases to the underlying gauge fields prior to source creation according to
\be
\tilde{U}_\mu\left[x\right]=e^{i\frac{2\pi}{L}\zeta d_\mu}U_\mu\left[x\right]
\ee
where $\vec{d}$ is the direction in which phases are applied, with magnitude $\zeta$ tuned for each desired momenta. The final interpolating fields are given by 
\bea
\Pi_{\vec{p}}(t) = \sum_{\vec{x}} e^{i\vec{p}\cdot\vec{x}} \bar{\tilde q}(\vec{x},t)\gamma^5 \tilde q(\vec{x},t) 
\eea
where $\tilde q$ is a light quark field constructed with the combined application of momentum smearing and Jacobi smearing. The smearing parameters used were varied for each momentum and are shown in Table~I. Due to
the decreasing signal to noise ratio, the higher momentum states required more source points and shorter time separation between the pion operators. These values are
also shown in Table~I.
\begin{table}[h!]
  \begin{center}
    \begin{tabular}{  c | c | c | c  }
      \hline
      \hline
      $\vec{p}=\left[0,0,p_z\right]$ & $\zeta$ & \vtop{\hbox{\strut No. of source points}\hbox{\strut $\qquad \quad (x_0,t)$}} &  \vtop{\hbox{\strut No. of source-sink}\hbox{\strut \quad\, separations}} \\ \hline
      $p=0.610\text{ GeV}$ & $1.75$ & $2$ & $9$\\
      $p=0.915\text{ GeV}$ & $2.50$ & $5$ & $9$\\
      $p=1.220\text{ GeV}$ & $3.75$ & $6$ & $9$\\
      $p=1.525\text{ GeV}$ & $4.50$ & $7$ & $7$\\ \hline \hline
    \end{tabular}
    \caption{The lattice momenta $\vec{p}=\left[0,0,p_z\right]$ of our interpolating operators and the momentum-smearing phases $\zeta$ applied for each lattice momenta in the direction $\vec{d}=\left[0,0,1\right]$, as well as number of pion source points and source-sink separations. Quark sources comprising our interpolators were subsequently spatially smeared according to the Jacobi-smearing procedure with smearing parameters $\sigma=4.0$ and $n_\sigma=50$.}
  \end{center}
  \label{tab:correlators}
\end{table}

For the calculation of any good LCS, the composite operators used have finite spatial extent $\xi$.   Introduction of a heavy auxiliary quark field $Q$ ($m_Q>m_l$), such that our operators are of the form $\mathcal{O}(t) = \mathcal{J}_{\Gamma}^\dagger(\xi,t) \mathcal{J}_{\Gamma'}(0,t)$ with $\mathcal{J}_{\Gamma} = \bar q \Gamma Q$, limits the available phase space between the two currents thereby reducing the statistical noise. An auxiliary heavy quark has also been used in Ref.~\cite{Detmold:2005gg} to remove higher twist contamination in the calculation of moments of the PDF and the distribution amplitude (DA). For our calculation of the pion valence distribution, multiple auxiliary quark masses between the light and strange quark mass were tested. A slight improvement in the signal-to-noise ratio from the heavier masses was observed for the larger momenta. We set the auxiliary quark propagator to the strange quark mass for the remainder of this calculation. In addition, to minimize excited state contamination, the operator insertion time ($t$) will be fixed to be midway between the source and sink interpolators ({\em i.e.} $t= \frac{T}{2}$). \\

\begin{figure}[t]
\begin{center}
\includegraphics[width=3.0in]{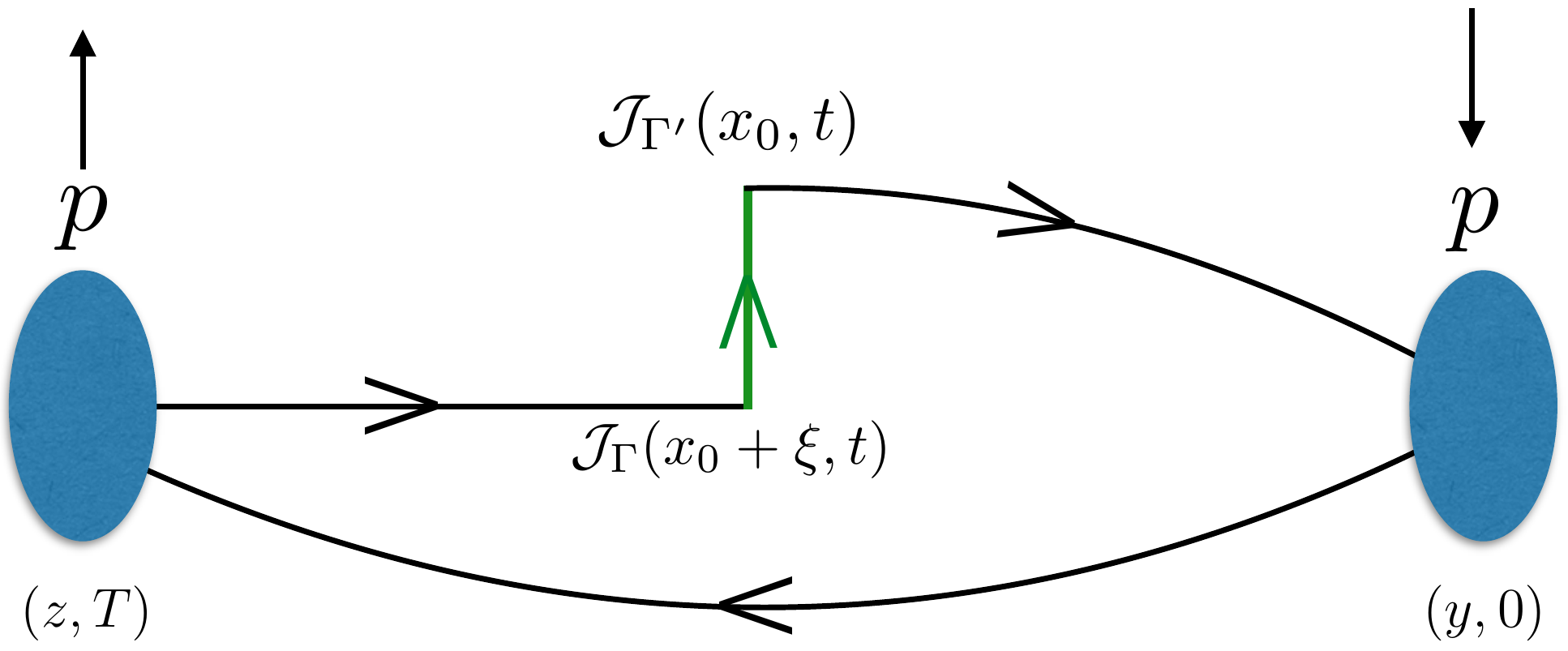}
\caption{\label{fig:4ptPic}
The 4pt-function is constructed by combining a heavy quark propagator, represented by the green line, and a modified sequential source, represented by the black lines. The sequential sources are made by determining  randomly chosen sample points $\left(x_0,t\right)$, inverting off each of these source points to the pion sources, and then using these objects for further inversions to the sinks and later to the second current locations. An advantage of this setup is that any pair of currents, $\mathcal{J}_\Gamma$ and $\mathcal{J}_\Gamma'$, with any separation $\xi$, may be constructed without additional costly propagator inversions.}
\end{center}
\end{figure}

The 4-point correlation function is constructed using a modified sequential source technique. Because we are not performing a time slice momentum projection at the operator, the standard sequential source method using the operator as sequential source does not work here. However, for the case of meson there is a straightforward implementation where momentum projections are performed at source and sink meson operators and the corresponding correlation functions are computed as chain of sequential sources as described below. The correlation function is expressed as follows
\bea
C_{\rm 4pt} \!\!\!\!\!\!\!&& (\xi,p,T,t)\nn \\
&=& \langle \Pi_p(\vec{z},T) \mathcal{J}_{\Gamma}^\dagger(x_0+\xi ,t) \mathcal{J}_{\Gamma'}(x_0,t)\overline{\Pi}_p\left(\vec{y},0\right)\rangle \nn \\
 &=&  \sum_{\vec{z},\vec{y}} e^{-i (\vec{z}-\vec{y})\cdot\vec{p}} \langle \bar{\tilde d}\gamma^5 \tilde u\left(\vec{z},T\right) ~ \bar Q\Gamma u(x_0+\xi ,t)\nn \\
 && \times\, \bar u\Gamma' Q(x_0 ,t) ~  \bar{\tilde u}\gamma^5 \tilde d\left(\vec{y},0\right)\rangle \nn \\
 &=& \Tr{\left[I_q^p(x_0+\xi,t; x_0, t) \Gamma \gamma^5 G_{Q}(x_0+\xi,t;x_0,t)^\dagger\gamma^5\Gamma'\right]}\nn \\
 \eea
 where $(x_0,t)$ is a randomly determined source point, $G_Q(y';x')$ is the flavor $Q$ auxiliary quark propagator from $x'$ to $y'$, and $I^p_q(y';x')$ is the modified sequential source with flavor $q$-quarks and pions at momentum $p$. The modified sequential source is constructed through sequential inversions of the light quark Dirac operator, reusing already calculated propagators. Heuristically, the modified sequential source is constructed by calculating the light quark propagator from a point-source located at one of the currents to the source interpolator, using this object as a source for a subsequent propagator to the sink interpolator, and lastly using this larger object as a source for propagation from the sink to the second current. This construction is done by solving the following sequence of systems of equations for $G_q$, $H^p_q$, and $I^p_q$.
 \bea \label{Seq-prop}
 \circled{1}&&\sum_{x',s'} D_q(x,s;x',s') G_q(x',s';x_0,t) = \delta(x-x_0)\delta(s-t) \nn \\
 \circled{2}&&\sum_{x',s'} D_q(x,s;x',s') H^p_q(x',s';x_0,t) =  e^{-ix\cdot p}\nn \\
 && \times \sum_{x',x''} S(x;x') \gamma^5 S(x';x'') G_q(x'',s;x_0,t) \delta(s)\nn  \\
 \circled{3}&&\sum_{x',s'} D_q(x,s;x',s') I^p_q(x',s';x_0,t) =  e^{ix\cdot p}\nn \\
 &&\times \sum_{x',x''} S(x;x') \gamma^5 S(x';x'') H^p_q(x'',s;x_0,t) \delta(s-T)\nn \\
 \eea
 where $D_q$ is the Dirac matrix for the light quarks and $S(x;x')$ represents the smearing procedure. The phase projects the interpolating fields onto definite momentum, while the signs of the momentum-smearing phases applied to the quark fields must be treated carefully to ensure $\Pi$ and $\overline{\Pi}$ correctly project onto states with a given momentum. Note that the smearing procedures are applied once for each of the quark fields in the interpolating fields, and no smearing is applied at the current insertions. This procedure is shown diagrammatically in Fig~\ref{fig:4ptPic}. An advantage of this procedure is that the correlation function for any current pair with any separation can be calculated without additional Dirac matrix inversions. On the other hand, for each choice of momentum and source-sink separations, two additional light quark Dirac matrix inversions are required per gauge configuration, denoted by \circled{1} and \circled{2} in Eq.~\eqref{Seq-prop}.
 
%
\section{Numerical Results and Extraction of Pion Valence Distribution} 
\label{results}

Reliable extraction of hadronic matrix elements, in part, hinges on how well a lattice calculation can systematically quantify and reduce excited-state contamination. In this section, we present the numerical results of our calculation of matrix elements of spatially-separated antisymmetric V-A currents.  As discussed in Sec.~\ref{methods},  the pion source-sink separation $T$ is systematically increased, while holding  the time $t=\frac{T}{2}$ fixed at which the currents are inserted.  Figure~\ref{matelem} shows the calculation of the operator $\bra{\pi(p)}\mathcal{J}_i(x_0+\xi)\mathcal{J}_j(x_0)\ket{\pi(p)}$, where $\mathcal{J}_i$ and $\mathcal{J}_j$ are the antisymmetric V-A currents discussed in Sec.~\ref{LOkernel}.

We perform a correlated fit to the jackknife ensemble ratios of 4pt to 2pt functions for a given momentum $p$ and spatial separation $\xi$ between the currents. In order to extract the desired matrix element from fits to our data, we assume the following single exponential form for the ratios of 4pt to 2pt functions:
\bea \label{fit}
R(T)= \frac{C_{\rm 4pt} (T)}{C_{\rm 2pt} (T)} = A + B e^{-\Delta_{\rm eff}T}
\eea
where  $\Delta_{\rm eff}$  is the effective energy gap between the ground-state and the excited states. Therefore the ratio $\frac{C_{\rm 4pt} (T)}{C_{\rm 2pt} (T)}$ will give the desired matrix element $\bra{\pi(p)}\mathcal{J}_i(x_0+\xi)\mathcal{J}_j(x_0)\ket{\pi(p)},$ up to an additional amplitude obtained from the fit to the 2pt function in the asymptotic limit of large $T$. Given the symmetries engineered into our calculation, namely the current always being inserted at $t=\frac{T}{2}$ and the source/sink interpolators being created in an identical manner, one can further assume that the excited state contamination is equal on the source and sink sides of the current. 

\begin{figure}[htp]
  \centering
  \subfigure[]{\includegraphics[width=0.4\textwidth]{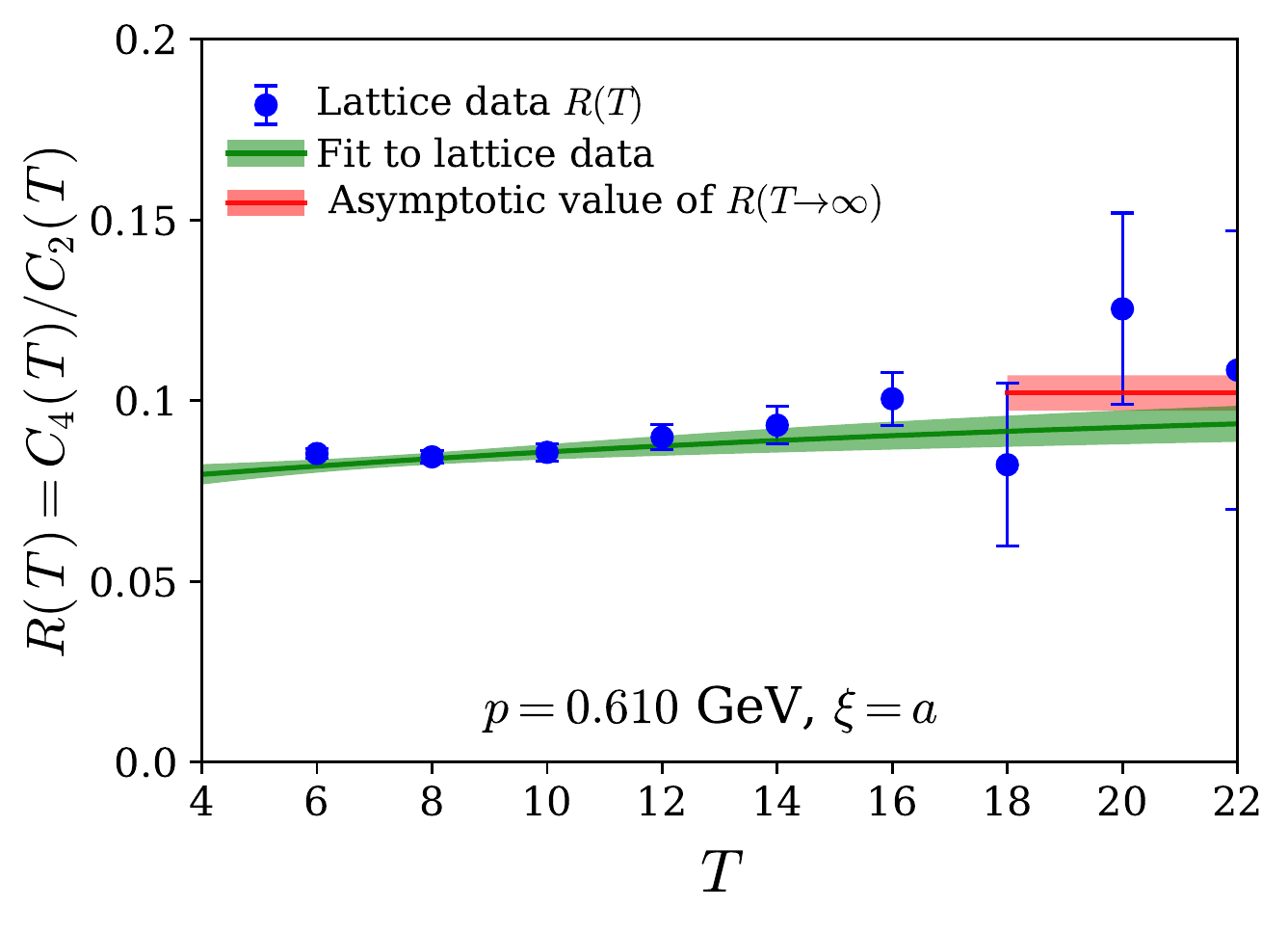}\label{2a}}
  \subfigure[]{\includegraphics[width=0.4\textwidth]{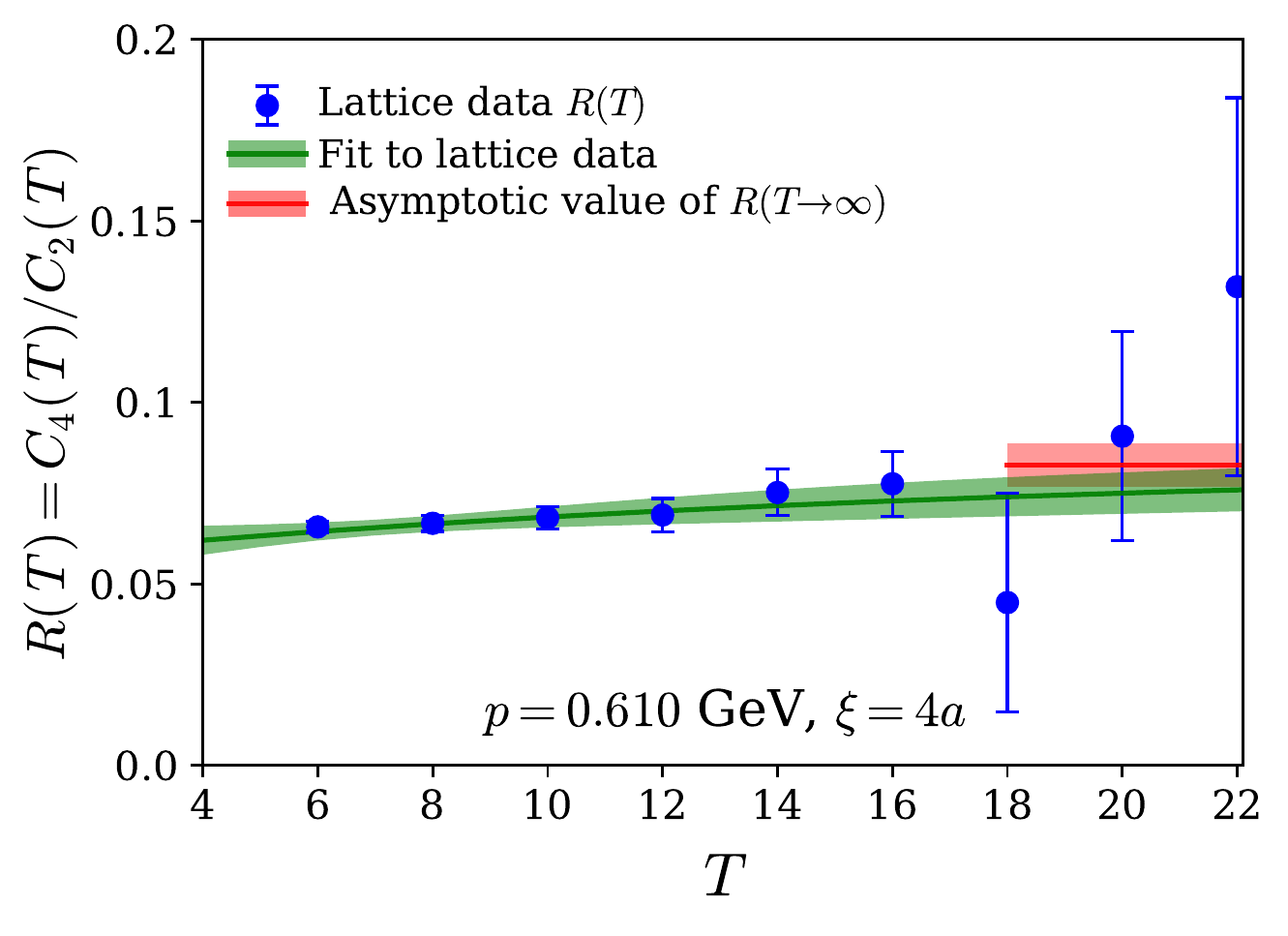}\label{2b}}
  \caption{\label{matelem}
   Jackknife ensemble ratio data of the 4pt to 2pt correlation functions used in the extraction of antisymmetric V-A current matrix elements. Figures~(\ref{2a}) and (\ref{2b}) show fits to the matrix elements for $p=0.610$ GeV with spatial separation between the currents $\xi=1a$ and $\xi=4a$, respectively.  The blue data points are obtained from the lattice QCD calculation, the green band shows the two-state fit to the data, and the red band shows the extracted value of the matrix element in the asymptotically large source-sink separation limit.  }
\end{figure}

In Figs.~(\ref{2a}) and ~(\ref{2b}) we present representative fits according to Eq.~\eqref{fit} applied to the jackknife ensemble ratio of 4pt ($C_{4{\rm pt}}$) to 2pt ($C_{2{\rm pt}}$) correlation functions as a function of source-sink separation $T$,  for $\vec{\xi}=\left[0,0,a\right]$ and momenta along the $z$-direction $p=0.610$ GeV and $p=1.525$ GeV, respectively. For $p=0.610$ GeV, the largest source-sink separation we use is $T=22a$. As the signal-to-noise ratio is significantly reduced for states with large momentum, we attempt to limit additional noise in our extracted matrix elements by considering smaller separations between the currents for states with large momenta. For instance,  we limit the largest source-sink separation to be $T=16a$ for $p=1.525$ GeV. As seen from Fig.~(\ref{2a}), the ratio of the correlation function for $p=0.610$ GeV has reasonable signal for almost all the source-sink separations. However, as expected for $p=1.525$ GeV, after $T=12a$, the lattice matrix  elements become very noisy and have no effect on the fit. Reasonable statistical signal in such a relatively small window of source-sink separations, compared to lower momenta data, might be a cause for concern. However it is worth remembering that with such a coarse lattice spacing ($a=0.127$ fm), $T=12a$ is sufficiently large ($\sim1.525$ fm) to minimize any excited-state contamination. We present values of the fit parameters in Table~\ref{table:r0}. 
\begin{table}[htbp]
\begin{center}
\begin{tabular}{c|c|c|c|c|c}
\hline \hline
$p$ [GeV] & $\xi$ & $A$  &$B$ &   $\Delta_{\rm eff}$  & $\chi^2$/d.o.f. \\
\hline
0.610 GeV & $1a$ & $0.102(5)$& -0.028(11) & 0.054(20)  & 1.21    \\
\hline
0.610 GeV &$4a$ &  $0.083(4)$& -0.026(13) & 0.062(42)  & 0.89    \\
\hline
1.525 GeV & $1a$ & $0.097(8)$& -0.267(513)    &0.809(503) & 0.15 \\
\hline
1.525 GeV & $4a$ & $0.049(9)$& -0.107(148)    &0.450(420) & 0.19 \\
\hline \hline
\end{tabular}
\end{center}
\caption{\label{table:r0} The fit parameters of the V-A matrix elements with $\vec{\xi}=\left[0,0,a\right]$ for momenta along the $z$-direction $p=0.610$ GeV and $p=1.525$ GeV, respectively. }
\end{table}

\begin{figure}[htp]
  \centering
  \subfigure[]{\includegraphics[width=0.4\textwidth]{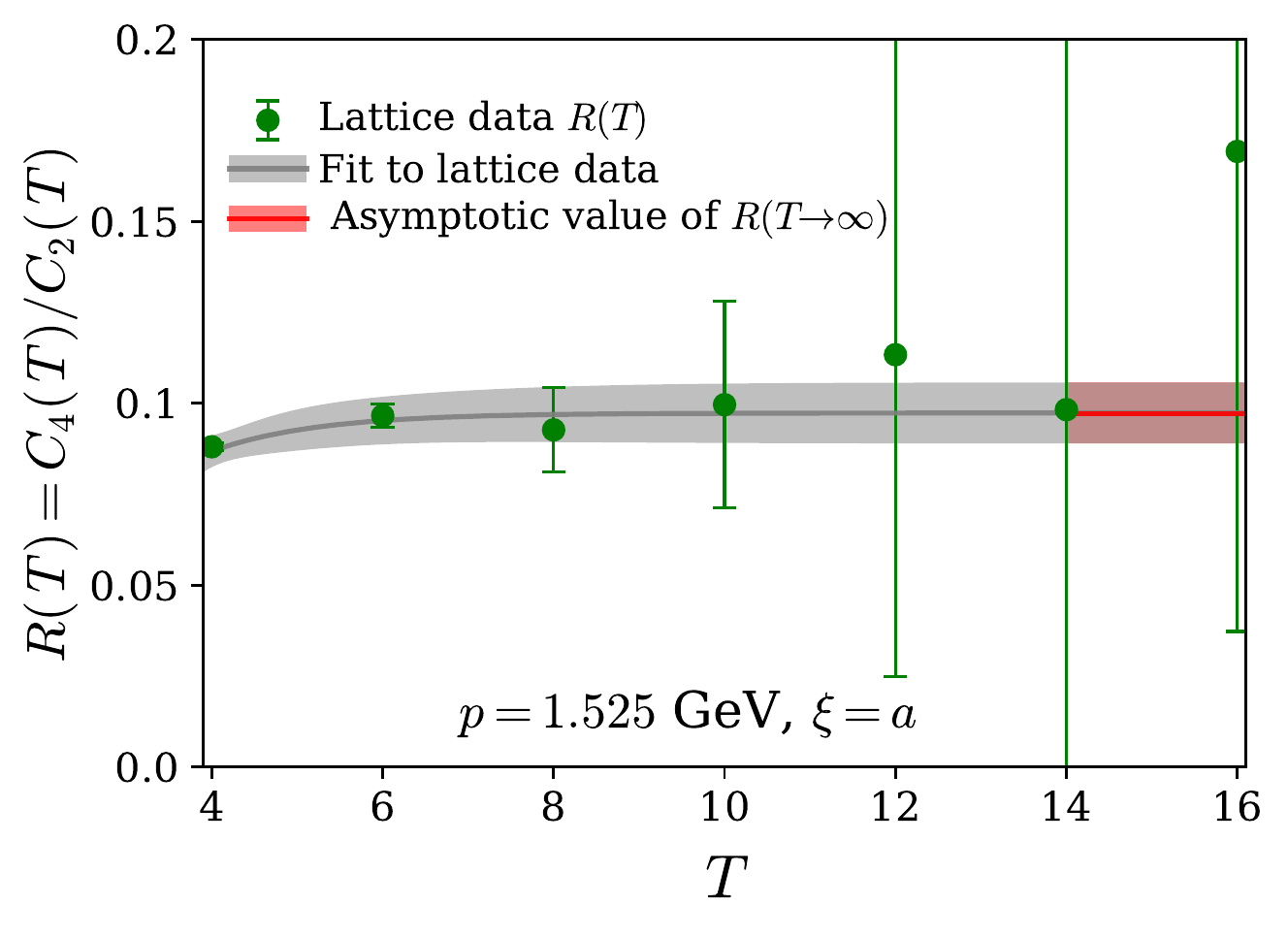}\label{3a}}
  \subfigure[]{\includegraphics[width=0.4\textwidth]{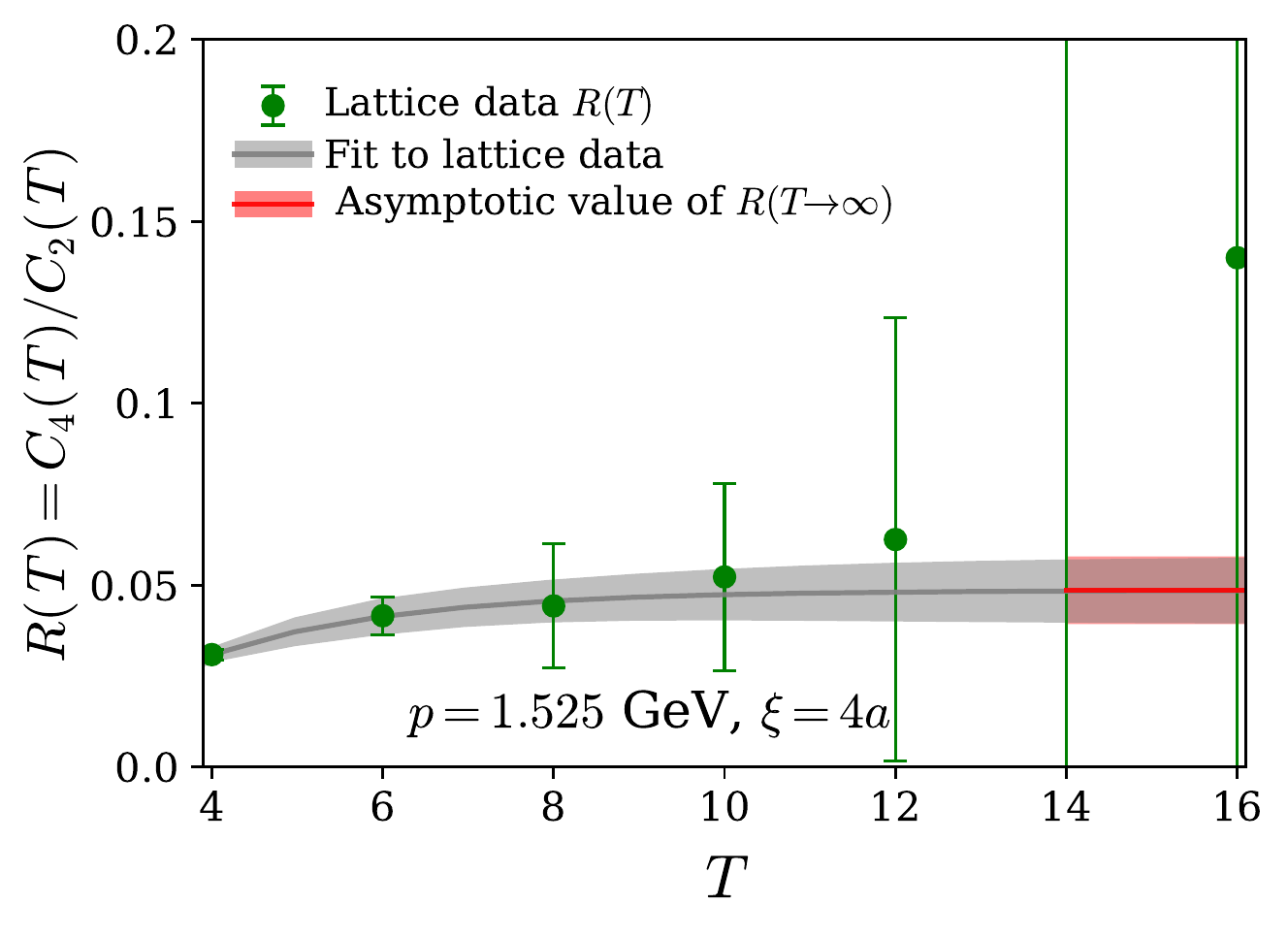}\label{3b}}
  \caption{\label{matelem}
   Jackknife ensemble ratio data of the 4pt to 2pt correlation functions used in the extraction of antisymmetric V-A current matrix elements. Figures~(\ref{3a}) and~(\ref{3b}) show fits to the matrix elements for $p=1.525$ GeV with spatial separation between the currents $\xi=1a$ and $\xi=4a$, respectively. The green data points are obtained from the lattice QCD calculation, the gray band shows the two-state fit to the data, and the red band shows the extracted value of the matrix element in the asymptotically large source-sink separation limit.  }
\end{figure}

\begin{figure}[htp]
\begin{center}
\includegraphics[width=3.4in]{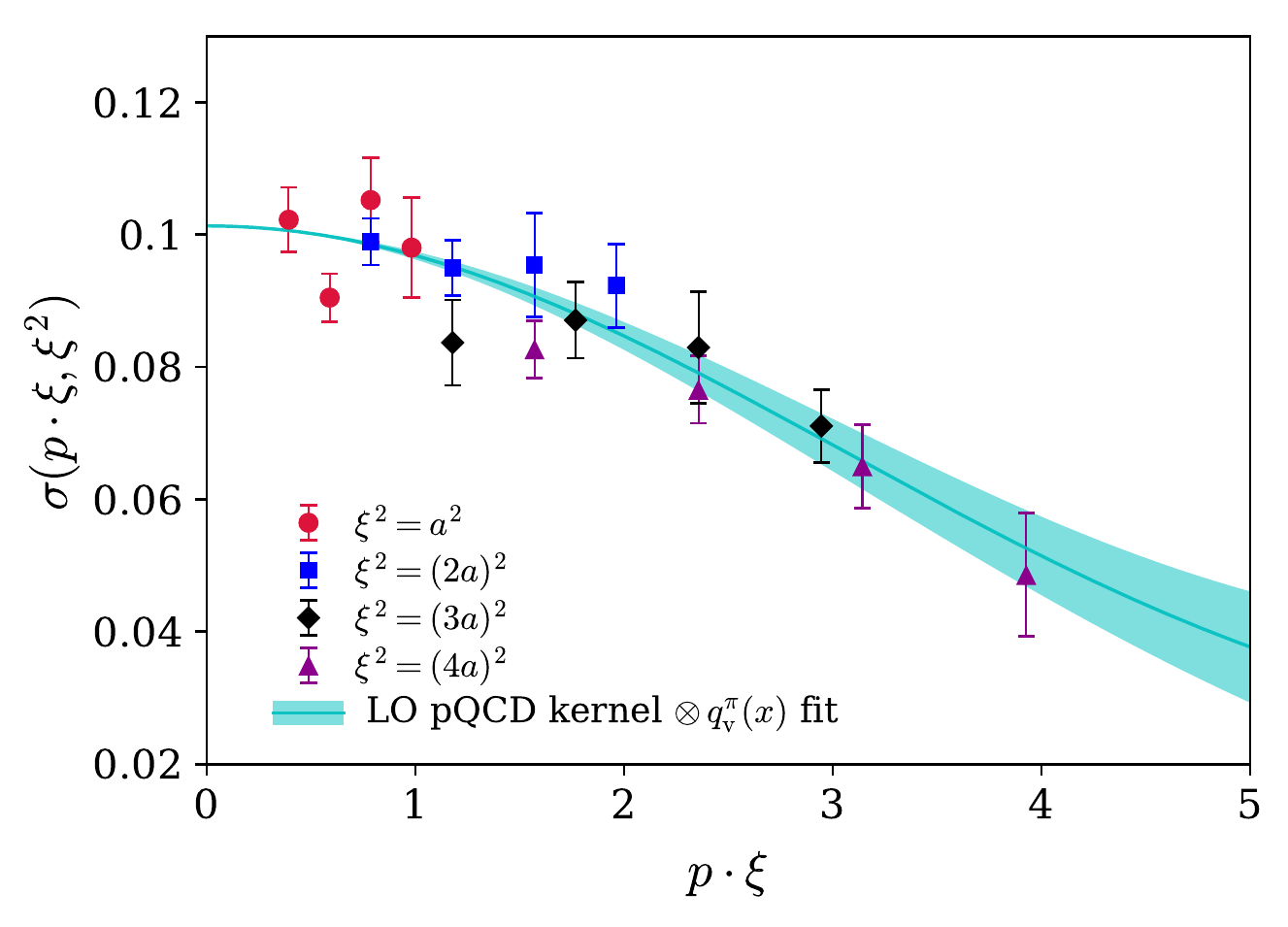}
\caption{\label{fig:LO}
Fit to the antisymmetric V-A currents matrix element with leading order (LO) perturbative kernel in Eq.~\eqref{pdf-fit} and functional form of pion valence distribution in Eq.~\eqref{pdf-form-use}. }
\end{center}
\end{figure}

With the fit  to the data for momenta along the $z$-direction in the range $p\in\lbrace0.610-1.525\rbrace$ GeV and current separations $\lvert\xi\rvert\le 4a$ in the $z$-direction, we obtain the matrix elements shown in Fig.~\ref{fig:LO}. As discussed in Sec.~\ref{LOkernel}, we only include $\lvert\xi\rvert\le 4a$ in our analysis so that $\xi$ is sufficiently smaller than $\frac{1}{\Lambda_{\rm QCD}}$ and thereby ensuring the factorization of Eq.~\eqref{eq:factor}.

For the lowest-order kernel, we use the following simple relation derived in Sec.~\ref{LOkernel}
\bea \label{pdf-fit}
T_1\left(\omega,\xi^2\right) \equiv& \sigma(p\cdot\xi, \xi^2)= \int_0^1 dx \frac{1}{\pi^2} \cos(x\omega)  q^\pi_{\rm v}(x)\nn \\
\eea
to extract pion valence distribution $q^\pi_{\rm v}(x)$, by fitting the antisymmetric V-A current matrix elements $\sigma(p\cdot\xi, \xi^2)$. 

For a proper extraction of the PDF and comparison to global fit results, one would need to extend the LO matching formula in Eq.~\eqref{pdf-fit} to include a higher order matching kernel between LCSs and PDFs. A next-to-leading-order (NLO) matching kernel would include $O(\alpha_s)$ logarithmic $\xi^2$ and constant corrections. The logarithmic terms contain the scale dependent DGLAP evolution. The constant terms contain the information on renormalization of the lattice QCD matrix  element and the partonic PDFs which leads to the scheme dependence.
There also can exist higher twist effects which contaminate the results without sufficiently small $\xi^2$. Finally there exist potential discretization errors from the small separation size, as well as rotational symmetry breaking effects as observed in~\cite{Bali:2018spj}. \\

The extraction of the PDF using Eq.~\eqref{pdf-fit} from lattice calculated data constitutes an ill-posed inverse problem. Lattice data will always be discretized and in a limited range of $\omega$. As demonstrated in~\cite{Karpie:2019eiq}, a na\"ive discretized inverse cosine transform would introduce numerical artifacts into the PDF. Solutions to this inverse problem require additional information or constraints. In the global fitting community, additional information is given in the form of smooth physically motivated functional forms as described below. PDFs extracted using this technique have been successfully shown to describe different physical processes, thereby assuring the universality of the nonperturbative PDFs. Of importance, it is known that the valence distributions of nucleon and pion are smooth functions of $x$ in the region $0<x<1$. In the spirit of the functional forms used in global fits of PDFs, we insert
\bea \label{pdf-form}
q^\pi_{\rm v}(x) = N x^\alpha(1-x)^\beta(1+ \rho \sqrt{x} + \gamma x)
\eea
into Eq.~\eqref{pdf-fit} and numerically perform the integration, where $N$ is the normalization such that
\bea \label{condition}
\int_0^1 dx\,q^\pi_{\rm v}(x) =1.
\eea

With the limitations related to $\xi^2$ corrections in mind, in this preliminary calculation, we use the numerical fitting program ROOT~\cite{ROOT} to fit bootstrap  samples of the V-A matrix elements and obtain a LO $q^\pi_{\rm v}(x)$-distribution.  The uncertainty band in the fit has been obtained from the fit results of the bootstrap samples. With the various sources of $\xi^2$ corrections not taken into account, we did not expect the matrix elements as a function of Ioffe time to fall upon a single curve -  consequently the $\chi^2/{\rm d.o.f.}$ was close to 2.2.  

\begin{figure}[htp]
  \centering
  \subfigure[]{\includegraphics[width=0.48\textwidth]{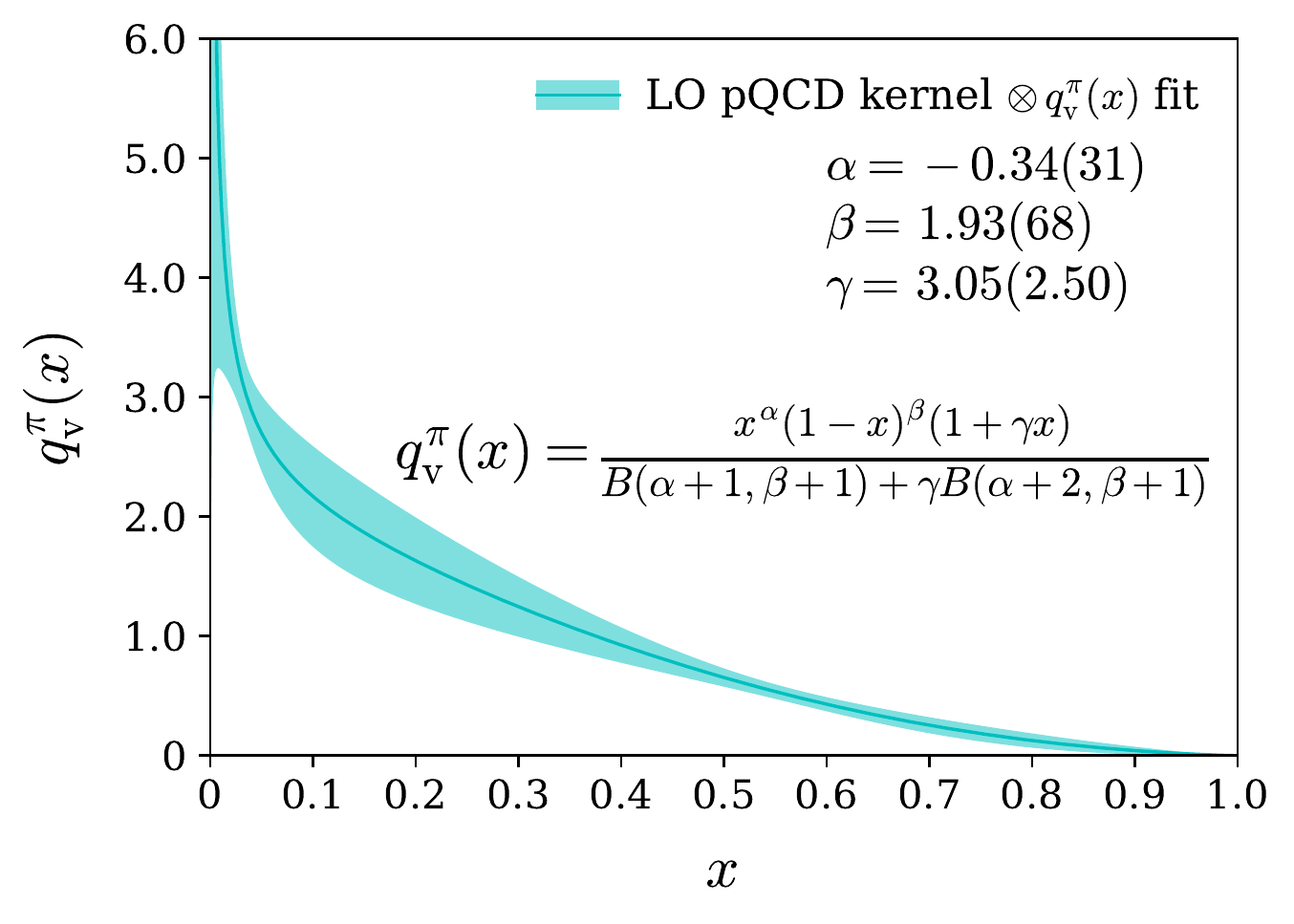}\label{4a}}
  \subfigure[]{\includegraphics[width=0.48\textwidth]{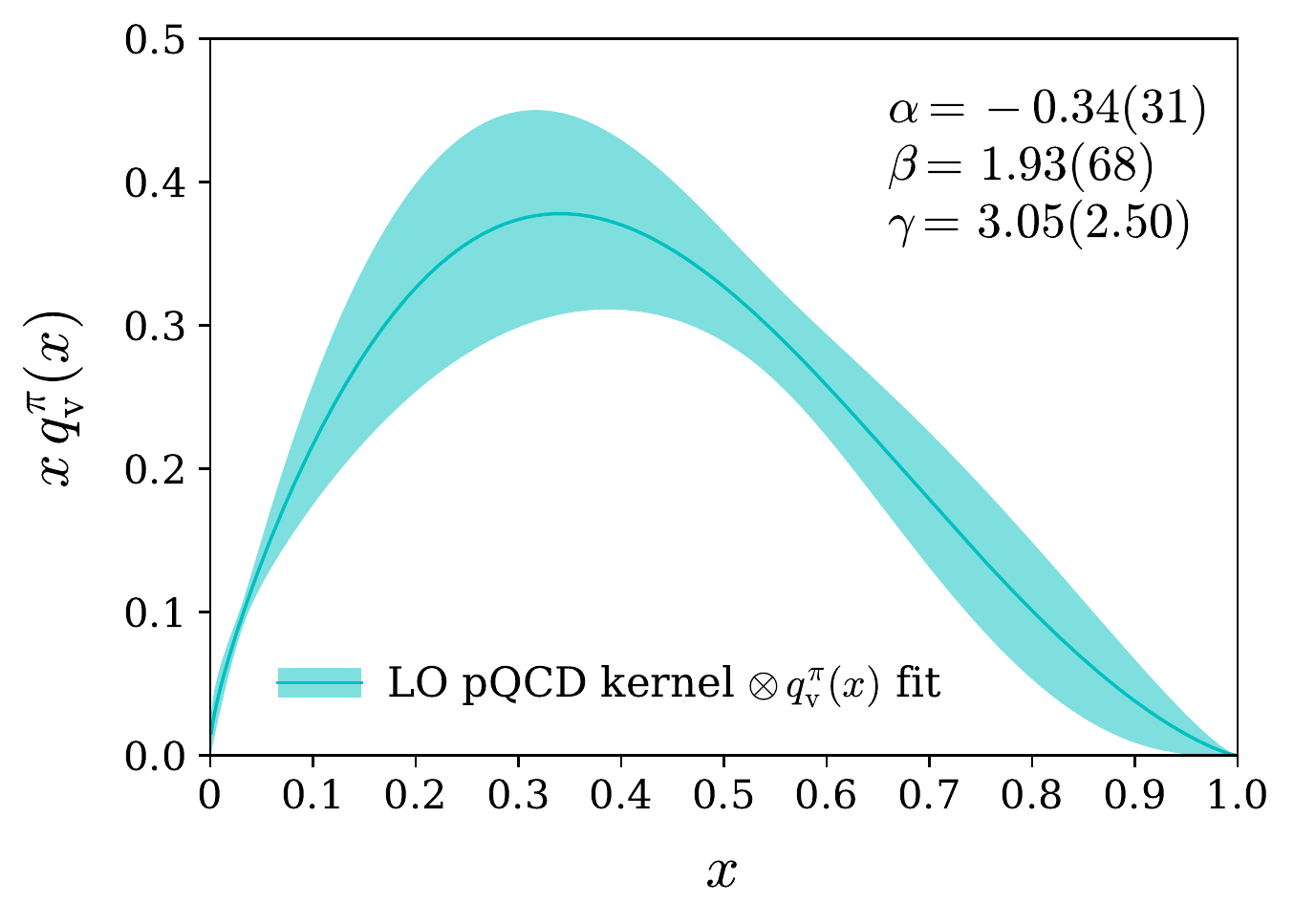}\label{4b}}
  \caption{\label{PDFres}
   The pion valence distribution obtained from the fit in Eq.~\eqref{pdf-fit} using the LO perturbative kernel in Eq.~\eqref{eq:C10} derived in Sec.~\ref{LOkernel} and the functional form of the PDF in Eq.~\eqref{pdf-form-use}. Figure~(\ref{4a}) shows the pion valence distribution $q^\pi_{\rm v}(x)$ and Figure~(\ref{4b})  shows the $xq^\pi_{\rm v}(x)$-distribution. The uncertainty band is obtained from the fits  to  the Jackknife samples of the data.}
\end{figure}

We find that the term $\rho \sqrt{x}$ in Eq.~\eqref{pdf-form} has no effect in the fit, as $\rho\simeq0$. A similar zero-value for $\rho$ was also found in Ref.~\cite{Chen:2016sno} and other global fits to experimental data. We therefore adopt a simpler functional form for the PDF in our calculation
\bea \label{pdf-form-use}
q^\pi_{\rm v}(x) = \frac{x^\alpha(1-x)^\beta(1+  \gamma x)}{B(\alpha+1,\beta+1)+\gamma B (\alpha+1+1, \beta+1)},
\eea
where the beta functions in the denominator ensure the normalization condition in Eq.~\eqref{condition} is met. In Eq.~\eqref{pdf-form-use}, the $(1-x)^\beta$ allows a smooth interpolation to zero as $x\to 1$ and is inspired by the counting rule of perturbative QCD. The $x^\alpha$ term is motivated by the behavior predicted by Regge theory at small $x$. One could interpolate these two limits using a polynomial of $x$. However, due to present statistics and small range of $\xi$, we cannot quantitatively distinguish between different choices of polynomials. Therefore,  we use the widely adopted phenomenologically motivated functional form of pion valence PDF in Eq.~\eqref{pdf-form}.  We set the following physically motivated and relaxed constraints
\bea
\alpha < 0, \quad  0< \beta < 4.
\eea
The fit to the lattice QCD data using the LO kernel in Eq.~\eqref{pdf-fit} and the functional form of PDF in Eq.~\eqref{pdf-form-use} is shown in Fig.~\ref{fig:LO} 
with the fit parameters,
\bea
\alpha &=& -0.34(31) \nn \\
\beta & = & 1.93(68) \nn \\
\gamma & = & 3.05 (2.50)
\eea 
The extracted PDF from this fit is shown in Fig.~(\ref{4a}) where the values of the fit parameters are indicated.  We also show the $xq^\pi_{\rm v}(x)$-distribution in Fig.~(\ref{4b}).  The perturbative kernel fixes the value of the integral in Eq.~\eqref{pdf-fit} to be $\frac{1}{\pi^2}$ at $\omega =0$ for any value of $x$, therefore the fitted value of $T_1\left(\omega,\xi^2\right)$ has zero uncertainty at this point. 


\section{Comparison with Other Determinations}

This first exploratory lattice QCD calculation of the pion PDF using spatially-separated current-current correlation function is performed at a relatively heavy pion mass ($m_\pi\simeq416\text{ MeV}$). This calculation must be repeated on several other lattice ensembles to determine the pion mass dependence, quantify lattice artifacts such as finite lattice spacing and finite volume~\cite{Briceno:2018lfj} corrections and obtain the PDF in the continuum limit. As mentioned earlier, extending the perturbative calculation beyond LO will not only lead to a more reliable extraction of the PDF, but also an understanding of power corrections and higher twist effects. A NLO matching kernel will give control over the corrections in $\xi$, both from DGLAP and higher twist effects. This calculation was performed on a fairly coarse lattice with a large minimum $\xi$, and in the future these corrections will need to be taken into account. While such calculations are underway and will be presented in a future work, the limitations in our current extraction of the pion valence PDF do not preclude comparison with global fits, two different model calculations and recent lattice calculations of pion valence quasi-distribution.

For a comparison with the LO extraction of $q^\pi_{\rm v}(x)$ from Drell-Yan experimental data in Ref.~\cite{Conway:1989fs}, we evolve our lattice QCD determination of the PDF in LO to an evolution scale of $\mu^2=27$ GeV$^2$ starting from initial scale of $\mu_0^2=1$ GeV$^2$. With only a LO matching kernel, the initial scale $\mu_0$ is chosen to be comparable to the $\frac{1}{\xi}$'s used in this calculation, but not low enough for perturbation theory to be doubted. With a NLO matching kernel, there will exist an explicit relationship between the scales $\xi$ and $\mu_0$ from the logarithmic terms. After the evolution, a shift in the peak of the $xq^\pi_{\rm v}(x)$-distribution toward smaller values of $x$ and a more convex-up behavior of the distribution near $x=1$ is seen as expected in our calculation.  From the fit parameters in Eq.~\eqref{pdf-form-use} ($\alpha=-0.34(31)$, $\beta=1.93(68)$, and $\gamma = 3.05(2.50)$ at the initial scale), it is seen that this lattice QCD calculation of $q^\pi_{\rm v}(x)$ is in agreement within uncertainty with the analysis in Ref.~\cite{Aicher:2010cb}, where the authors included next-to-leading-logarithmic threshold soft-gluon resummation effects in the calculation of the Drell-Yan cross section. The large-$x$ behavior is statistically consistent with the expectation based on perturbative QCD~\cite{Farrar:1979aw,Berger:1979du,Brodsky:1994kg} but of course with large uncertainty. In contrast, the large-$x$ behavior of this calculation has about $\sim \!\!1\sigma$ difference from the two other NLO fits~\cite{Wijesooriya:2005ir,Barry:2018ort} which obtained a harder $(1-x)$ fall-off of the pion valence distribution. 

\begin{figure}[htp]
\begin{center}
\includegraphics[width=3.5in]{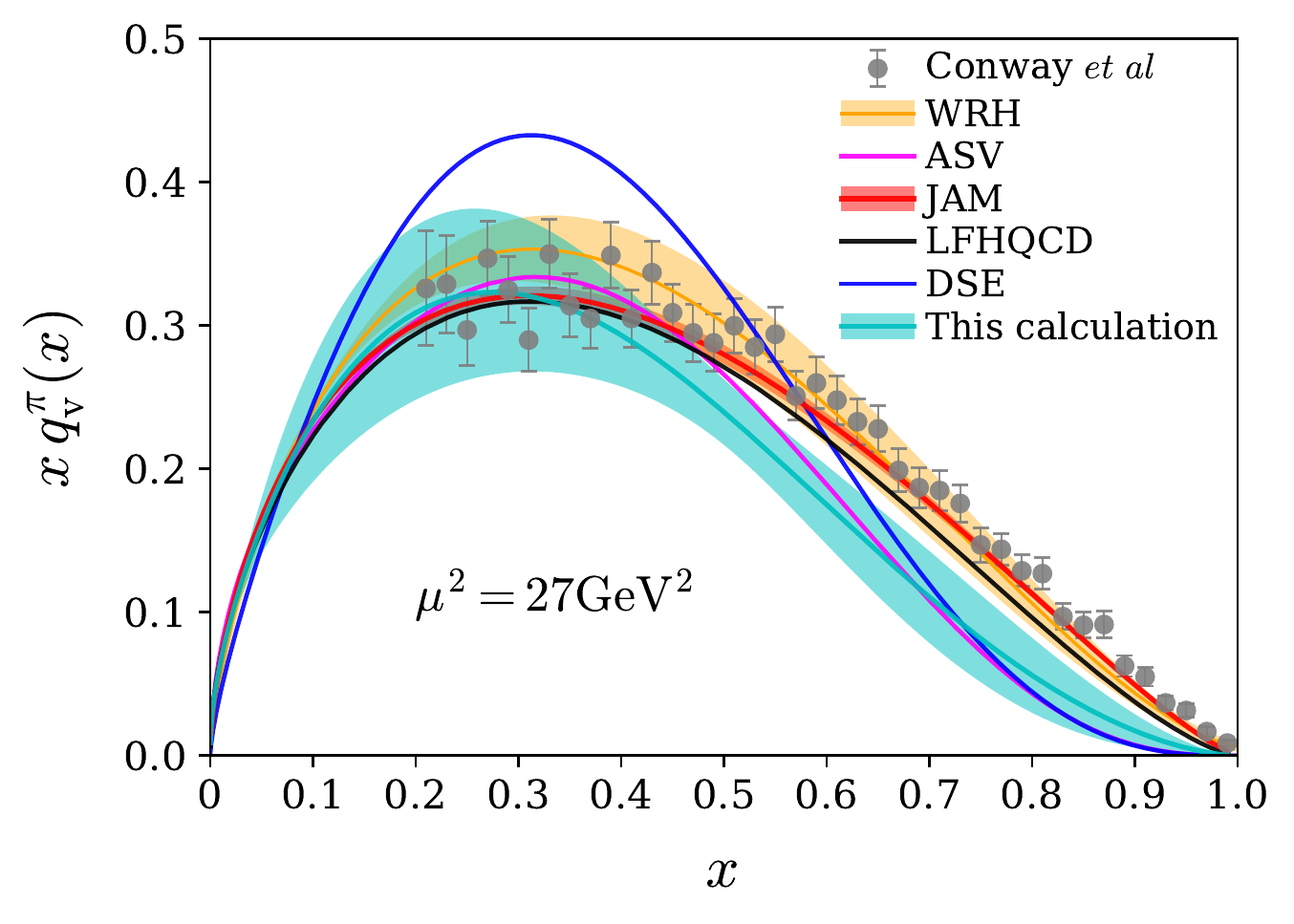}
\caption{\label{fig:evolPDF}
Comparison of pion $xq^\pi_{\rm v}(x)$-distribution with the leading-order (LO) extraction from Drell-Yan data~\cite{Conway:1989fs} (gray data points with uncertainties), next-to-leading order (NLO) fits~\cite{Wijesooriya:2005ir,Aicher:2010cb,Barry:2018ort} (orange band, magenta curve, and red band),  and model calculations~\cite{deTeramond:2018ecg,Chen:2016sno} (black and blue lines). This lattice QCD calculation of $q^\pi_{\rm v}(x)$ is evolved from an initial scale $\mu_0^2=1$ GeV$^2$ at LO. All the results are at evolved to an evolution scale of $\mu^2=27$ GeV$^2$. }
\end{center}
\end{figure}

It is seen in Fig.~(\ref{fig:evolPDF}) that the large-$x$ behavior of this calculation is statistically consistent with the Dyson-Schwinger model prediction~\cite{Chen:2016sno} labeled as ``DSE'' in the momentum fraction region $x> 0.7$. On the other hand, this lattice QCD calculation of $q^\pi_{\rm v}(x)$ is in statistical agreement with the light-front holographic QCD model calculation labeled as ``LFHQCD" in the region $x<0.5$, but shows a slightly softer fall-off at large-$x$ in its central value. As mentioned earlier, in a future calculation, when all the systematics of this lattice QCD calculation are to be well understood and controlled in a proper way, the first-principles determination of large-$x$ behavior of pion PDF such as this one can shed light for understanding different approximations used in an array of model calculations. 

Even with the limitations mentioned above, this first exploratory lattice QCD determination of  $q^\pi_{\rm v}(x)$ using LCSs provides encouraging results  and shows that this method has the potential to capture the essential dynamics dictating the behavior of hadron PDFs. Of notable interest, our calculation of $x q^\pi_{\rm v}(x)$ , shown in Figs.~\ref{PDFres} and~\ref{fig:evolPDF}, illustrates a peak of the distribution in a region $x<0.50$ at any scale $\mu^2$. This is consistent with all the global analyses of the pion valence distribution, wherein $x q^\pi_{\rm v}(x)$ is peaked below $x=0.50$. The readers are referred to~\cite{Chen:2018fwa} and \cite{Karthik:2018wmj} for recent other lattice calculations of the pion valence quasi-distribution.


\section{Summary and Outlook}

We have presented the first lattice QCD calculation of the pion valence distribution using a spatially-separated vector and axial-vector current combination. We have emphasized that the spatial separation $\xi$ between the currents is a well-defined quantity in the good lattice cross sections method and plays a role analogous to capturing the correct collision dynamics in a hard scattering process and ensures the validity of factorization to obtain parton distribution functions. In this exploratory calculation, we have considered a leading-order perturbative kernel to obtain the nonperturbative valence PDF of the pion though factorization of the good lattice cross section of this vector and axial-vector matrix element. A similar calculation on other lattice ensembles is in progress to determine the pion mass dependence, quantify lattice artifacts and obtain the PDF in the continuum limit. Such a calculation of the lattice QCD matrix elements in the continuum limit, and therefore a more reliable extraction of PDF will be presented in future work with the next-to-leading-order perturbative matching kernel incorporated to understand the corrections in $\xi$ and higher twist effects. Moreover, using this most general approach, other good lattice cross sections with different current combinations will give information on different types of PDFs. Within the limitations of the present calculation, however, we would like to emphasize that the good statistical agreement between the PDF extracted here, through only leading-order factorization and the fits to the experimental data, is very encouraging and shows that this method has the potential to complement the well-established and modern state-of-art global fits of PDFs.  Upon further investigation and refinement of our methodology,  our lattice QCD results can be a subset of those used in future global analyses.
\section*{Acknowledgments}
R.S.S., J.K., and C.E. give special thanks to Robert G. Edwards whose help has been tremendous to set up the numerical calculation.  We thank Ra\'ul A. Brice\~no, Martha Constantinou, Nikhil Karthik, Luka Leskovec, Tianbo Liu, Ying-Qing Ma, Nobuo Sato, Yi-Bo Yang, Jian-Hui Zhang who provided insights and expertise that greatly assisted this research. \\

This work is supported by the U.S. Department of Energy, Office of Science, Office of Nuclear Physics under Contract No. DE-AC05-06OR23177, within the framework of the TMD Collaboration.  We acknowledge the facilities of the USQCD Collaboration
used for this research in part, which are funded by the Office of Science of the U.S. Department
of Energy. This research used resources of the Oak Ridge Leadership Computing Facility at the Oak Ridge National Laboratory, which is supported by the Office of Science of the U.S. Department of Energy under Contract No. DE-AC05-00OR22725. J.K. and C.E. are supported in part by the U.S. Department of Energy under contract DE-FG02-04ER41302 and Department of Energy Office of Science Graduate Student Research fellowships, through the U.S. Department of Energy, Office of Science, Office of Workforce Development for Teachers and Scientists, Office of Science Graduate Student Research (SCGSR) program. K.O. acknowledges support in part  by the U.S. Department of Energy through Grant Number DE- FG02-04ER41302, by STFC consolidated grant ST/P000681/1, and the hospitality from DAMTP and Clare Hall at Cambridge University.

\providecommand{\href}[2]{#2}
\begingroup\raggedright

\endgroup


\begin{thebibliography}{10}



\bibitem{Collins:1989gx} 
  J.~C.~Collins, D.~E.~Soper and G.~F.~Sterman,
  ``Factorization of Hard Processes in QCD,''
  \href{https://doi.org/10.1142/9789814503266_0001}{Adv.\ Ser.\ Direct.\ High Energy Phys.\  {\bf 5}, 1 (1989)}
  [\href{https://arxiv.org/abs/hep-ph/0409313}{\tt  arXiv:hep-ph/0409313}].
  
  
 \bibitem{Feynman} 
  R.~P.~Feynman, ``Photon-hadron interactions," 
  W. A. Benjamin, Jan 1, Reading 1972 - Hadrons - 282 p



  
  
  \bibitem{Harland-Lang:2014zoa} 
  L.~A.~Harland-Lang, A.~D.~Martin, P.~Motylinski and R.~S.~Thorne,
  Parton distributions in the LHC era: MMHT 2014 PDFs,
  \href{https://doi.org/10.1140/epjc/s10052-015-3397-6}{Eur.\ Phys.\ J.\ C {\bf 75}, 204 (2015)}
  [\href{https://arxiv.org/abs/1412.3989}{\tt arXiv:1412.3989 [hep-ph]}].
  
 
\bibitem{Dulat:2015mca} 
  S.~Dulat {\it et al.},
  New parton distribution functions from a global analysis of quantum chromodynamics,
  \href{https://doi.org/10.1103/PhysRevD.93.033006}{Phys.\ Rev.\ D {\bf 93},  033006 (2016)}
   [\href{https://arxiv.org/abs/1506.07443}{\tt arXiv:1506.07443 [hep-ph]}].
  
 
 \bibitem{Ball:2017nwa} 
  R.~D.~Ball {\it et al.} (NNPDF Collaboration),
  Parton distributions from high-precision collider data,
  \href{https://doi.org/10.1140/epjc/s10052-017-5199-5}{Eur.\ Phys.\ J.\ C {\bf 77},  663 (2017)}
  [\href{https://arxiv.org/abs/1706.00428}{\tt arXiv:1706.00428 [hep-ph]}].
  
  
   \bibitem{Alekhin:2017kpj} 
 S.~Alekhin, J.~Bl\"umlein, S.~Moch and R.~Pla\v{c}akyt\.e,
 Parton distribution functions, $\alpha_s$, and heavy-quark masses for LHC Run II,
 \href{https://doi.org/10.1103/PhysRevD.96.014011}{Phys.\ Rev.\ D {\bf 96}, 014011 (2017)}
  [\href{https://arxiv.org/abs/1701.05838}{\tt arXiv:1701.05838 [hep-ph]}].
  
  
\bibitem{Ethier:2017zbq} 
  J.~J.~Ethier, N.~Sato and W.~Melnitchouk,
  ``First simultaneous extraction of spin-dependent parton distributions and fragmentation functions from a global QCD analysis,''
  \href{https://doi.org/10.1103/PhysRevLett.119.132001}{Phys.\ Rev.\ Lett.\  {\bf 119}, no. 13, 132001 (2017)}
  [\href{https://arxiv.org/abs/1705.05889}{\tt arXiv:1705.05889 [hep-ph]}].
  
  
    
\bibitem{Badier:1983mj} 
  J.~Badier {\it et al.} [NA3 Collaboration],
  ``Experimental Determination of the pi Meson Structure Functions by the Drell-Yan Mechanism,''
  \href{https://doi.org/10.1007/BF01573728}{Z.\ Phys.\ C {\bf 18}, 281 (1983)}.
  
  
\bibitem{Betev:1985pf} 
  B.~Betev {\it et al.} [NA10 Collaboration],
  ``Differential Cross-section of High Mass Muon Pairs Produced by a 194-{GeV}/$c \pi^-$ Beam on a Tungsten Target,''
  \href{https://doi.org/10.1007/BF01550243}{Z.\ Phys.\ C {\bf 28}, 9 (1985)}.
  
  
\bibitem{Conway:1989fs} 
  J.~S.~Conway {\it et al.},
  ``Experimental Study of Muon Pairs Produced by 252-GeV Pions on Tungsten,''
  \href{https://doi.org/10.1103/PhysRevD.39.92}{Phys.\ Rev.\ D {\bf 39}, 92 (1989)}.
  
  
\bibitem{Owens:1984zj} 
  J.~F.~Owens,
  ``$Q^2$ Dependent Parametrizations of Pion Parton Distribution Functions,''
  \href{https://doi.org/10.1103/PhysRevD.30.943}{Phys.\ Rev.\ D {\bf 30}, 943 (1984)}.
  
  
  \bibitem{Aurenche:1989sx} 
  P.~Aurenche, R.~Baier, M.~Fontannaz, M.~N.~Kienzle-Focacci and M.~Werlen,
  ``The Gluon Content of the Pion From High $p_T$ Direct Photon Production,''
  \href{https://doi.org/10.1016/0370-2693(89)91351-8}{Phys.\ Lett.\ B {\bf 233}, 517 (1989)}.
  
  \bibitem{Sutton:1991ay} 
  P.~J.~Sutton, A.~D.~Martin, R.~G.~Roberts and W.~J.~Stirling,
  ``Parton distributions for the pion extracted from Drell-Yan and prompt photon experiments,''
  \href{https://doi.org/10.1103/PhysRevD.45.2349}{Phys.\ Rev.\ D {\bf 45}, 2349 (1992)}.
  
  \bibitem{Gluck:1991ey} 
  M.~Gluck, E.~Reya and A.~Vogt,
  ``Pionic parton distributions,''
 \href{https://link.springer.com/article/10.1007%2FBF01559743}{ Z.\ Phys.\ C {\bf 53}, 651 (1992)}.
 
 \bibitem{Wijesooriya:2005ir} 
  K.~Wijesooriya, P.~E.~Reimer and R.~J.~Holt,
  ``The pion parton distribution function in the valence region,''
  \href{https://doi.org/10.1103/PhysRevC.72.065203}{Phys.\ Rev.\ C {\bf 72}, 065203 (2005)}
  [\href{https://arxiv.org/abs/nucl-ex/0509012}{\tt arXiv:nucl-ex/0509012}].
  
  \bibitem{Aicher:2010cb} 
  M.~Aicher, A.~Schafer and W.~Vogelsang,
  ``Soft-gluon resummation and the valence parton distribution function of the pion,''
  \href{https://doi.org/10.1103/PhysRevLett.105.252003}{Phys.\ Rev.\ Lett.\  {\bf 105}, 252003 (2010)}
  [\href{https://arxiv.org/abs/1009.2481}{\tt arXiv:1009.2481 [hep-ph]}].
  
  \bibitem{Barry:2018ort} 
  P.~C.~Barry, N.~Sato, W.~Melnitchouk and C.~R.~Ji,
  ``First Monte Carlo Global QCD Analysis of Pion Parton Distributions,''
  \href{https://doi.org/10.1103/PhysRevLett.121.152001}{Phys.\ Rev.\ Lett.\  {\bf 121}, no. 15, 152001 (2018)}
  [\href{arXiv:1804.01965 [hep-ph]}{\tt arXiv:1804.01965 [hep-ph]}].
  
  
  \bibitem{Farrar:1979aw} 
  G.~R.~Farrar and D.~R.~Jackson,
  ``The Pion Form-Factor,''
  \href{https://doi.org/10.1103/PhysRevLett.43.246}{Phys.\ Rev.\ Lett.\  {\bf 43}, 246 (1979)}.
  
  \bibitem{Berger:1979du} 
  E.~L.~Berger and S.~J.~Brodsky,
  ``Quark Structure Functions of Mesons and the Drell-Yan Process,''
  \href{https://doi.org/10.1103/PhysRevLett.42.940}{Phys.\ Rev.\ Lett.\  {\bf 42}, 940 (1979)}.
  
  \bibitem{Brodsky:1994kg} 
  S.~J.~Brodsky, M.~Burkardt and I.~Schmidt,
  ``Perturbative QCD constraints on the shape of polarized quark and gluon distributions,''
  \href{https://doi.org/10.1016/0550-3213(95)00009-H}{Nucl.\ Phys.\ B {\bf 441}, 197 (1995)}
  [\href{https://arxiv.org/abs/hep-ph/9401328}{\tt arXiv:hep-ph/9401328}].
  
\bibitem{Shigetani:1993dx} 
  T.~Shigetani, K.~Suzuki and H.~Toki,
  ``Pion structure function in the Nambu and Jona-Lasinio model,''
  \href{https://doi.org/10.1016/0370-2693(93)91302-4}{Phys.\ Lett.\ B {\bf 308}, 383 (1993)}
  [\href{https://arxiv.org/abs/hep-ph/9402286}{\tt arXiv:hep-ph/9402286}].
  
  
  \bibitem{Davidson:1994uv} 
  R.~M.~Davidson and E.~Ruiz Arriola,
  ``Structure functions of pseudoscalar mesons in the SU(3) NJL model,''
  \href{https://doi.org/10.1016/0370-2693(95)00091-X}{Phys.\ Lett.\ B {\bf 348}, 163 (1995)}.
  
\bibitem{Melnitchouk:2002gh} 
  W.~Melnitchouk,
  ``Quark hadron duality in electron pion scattering,''
  \href{https://doi.org/10.1140/epja/i2003-10006-6}{Eur.\ Phys.\ J.\ A {\bf 17}, 223 (2003)}
  [\href{https://arxiv.org/abs/hep-ph/0208258}{\tt arXiv:hep-ph/0208258}].
  
  \bibitem{deTeramond:2018ecg} 
  G.~F.~de T\'eramond, T. Liu,  R.~S.~Sufian, H.~G.~Dosch, S.~J.~Brodsky and A.~Deur[HLFHS Collaboration],
  ``Universality of Generalized Parton Distributions in Light-Front Holographic QCD,''
  \href{https://doi.org/10.1103/PhysRevLett.120.182001}{Phys.\ Rev.\ Lett.\  {\bf 120}, no. 18, 182001 (2018)}
  [\href{https://arxiv.org/abs/1801.09154}{\tt arXiv:1801.09154 [hep-ph]}].
  
  
  \bibitem{Hecht:2000xa} 
  M.~B.~Hecht, C.~D.~Roberts and S.~M.~Schmidt,
  ``Valence quark distributions in the pion,''
  \href{https://doi.org/10.1103/PhysRevC.63.025213}{Phys.\ Rev.\ C {\bf 63}, 025213 (2001)}
  [\href{https://arxiv.org/abs/nucl-th/0008049}{\tt arXiv:nucl-th/0008049}].
  
  
  \bibitem{Chen:2016sno} 
  C.~Chen, L.~Chang, C.~D.~Roberts, S.~Wan and H.~S.~Zong,
  Valence-quark distribution functions in the kaon and pion,
  \href{https://doi.org/10.1103/PhysRevD.93.074021}{Phys.\ Rev.\ D {\bf 93},  074021 (2016)}
  [\href{https://arxiv.org/abs/1602.01502}{\tt arXiv:1602.01502 [nucl-th]}].  
  
  
  
  
  
\bibitem{Ma:2014jla} 
  Y.~Q.~Ma and J.~W.~Qiu,
  ``Extracting Parton Distribution Functions from Lattice QCD Calculations,''
  \href{https://doi.org/10.1103/PhysRevD.98.074021}{Phys.\ Rev.\ D {\bf 98}, no. 7, 074021 (2018)}
  [\href{https://arxiv.org/abs/1404.6860}{\tt arXiv:1404.6860 [hep-ph]}].


  
  
  \bibitem{Ma:2017pxb} 
  Y.~Q.~Ma and J.~W.~Qiu,
  ``Exploring Partonic Structure of Hadrons Using ab initio Lattice QCD Calculations,''
  \href{https://doi.org/10.1103/PhysRevLett.120.022003}{Phys.\ Rev.\ Lett.\  {\bf 120}, no. 2, 022003 (2018)}
  [\href{https://arxiv.org/abs/1709.03018}{\tt arXiv:1709.03018 [hep-ph]}].
  
  
  


  
\bibitem{Feynman:1969ej} 
  R.~P.~Feynman,
  ``Very high-energy collisions of hadrons,''
  \href{https://doi.org/10.1103/PhysRevLett.23.1415}{Phys.\ Rev.\ Lett.\  {\bf 23}, 1415 (1969).}
  
  
   \bibitem{Liu:1993cv} 
  K.~F.~Liu and S.~J.~Dong,
  Origin of difference between $\overline d$ and $\overline u$ partons in the nucleon,
  \href{https://doi.org/10.1103/PhysRevLett.72.1790}{Phys.\ Rev.\ Lett.\  {\bf 72}, 1790 (1994)}
  [\href{https://arxiv.org/abs/hep-ph/9306299}{\tt arXiv:hep-ph/9306299}].
  
  
\bibitem{Liu:1999ak} 
  K.~F.~Liu,
  Parton degrees of freedom from the path-integral formalism,
  \href{https://doi.org/10.1103/PhysRevD.62.074501}{Phys.\ Rev.\ D {\bf 62}, 074501 (2000)}
  [\href{https://arxiv.org/abs/hep-ph/9910306}{\tt arXiv:hep-ph/9910306}].
  
 

  
  
  
\bibitem{Horsley:2012pz} 
  R.~Horsley {\it et al.} (QCDSF-UKQCD Collaboration),
  A lattice study of the glue in the nucleon,
  \href{https://doi.org/10.1016/j.physletb.2012.07.004}{Phys.\ Lett.\ B {\bf 714}, 312 (2012)}
  [\href{https://arxiv.org/abs/1205.6410}{\tt arXiv:1205.6410 [hep-lat]}].
    
    

  
  
\bibitem{Ji:2013dva} 
  X.~Ji,
  Parton physics on a Euclidean lattice,
  \href{https://doi.org/10.1103/PhysRevLett.110.262002}{Phys.\ Rev.\ Lett.\  {\bf 110}, 262002 (2013)}
  [\href{https://arxiv.org/abs/1305.1539}{\tt arXiv:1305.1539 [hep-ph]}].
  
  
  \bibitem{Radyushkin:2017cyf} 
  A.~V.~Radyushkin,
  Quasi-parton distribution functions, momentum distributions, and pseudo-parton distribution functions,
  \href{https://doi.org/10.1103/PhysRevD.96.034025}{Phys.\ Rev.\ D {\bf 96},  034025 (2017)}
  [\href{https://arxiv.org/abs/1705.01488}{\tt arXiv:1705.01488 [hep-ph]}].
  
  
    \bibitem{Braun:2007wv} 
  V.~Braun and D.~Mueller,
  ``Exclusive processes in position space and the pion distribution amplitude,''
  \href{https://doi.org/10.1140/epjc/s10052-008-0608-4}{Eur.\ Phys.\ J.\ C {\bf 55}, 349 (2008)}
  [\href{https://arxiv.org/abs/0709.1348}{\tt arXiv:0709.1348 [hep-ph]}].
  
  
 \bibitem{Bali:2018spj} 
  G.~S.~Bali {\it et al.},
  ``Pion distribution amplitude from Euclidean correlation functions: Exploring universality and higher-twist effects,''
  \href{https://doi.org/10.1103/PhysRevD.98.094507}{Phys.\ Rev.\ D {\bf 98}, no. 9, 094507 (2018)}
  [\href{https://arxiv.org/abs/1807.06671}{\tt arXiv:1807.06671 [hep-lat]}].

  
  \bibitem{Chambers:2017dov} 
  A.~J.~Chambers {\it et al.},
  Nucleon structure functions from operator product expansion on the lattice,
  \href{https://doi.org/10.1103/PhysRevLett.118.242001}{Phys.\ Rev.\ Lett.\  {\bf 118}, 242001 (2017)}
  [\href{https://arxiv.org/abs/1703.01153}{\tt arXiv:1703.01153 [hep-lat]}]. 
  
  
  \bibitem{Alexandrou:2018pbm} 
  C.~Alexandrou, K.~Cichy, M.~Constantinou, K.~Jansen, A.~Scapellato and F.~Steffens,
  ``Light-Cone Parton Distribution Functions from Lattice QCD,''
  \href{https://doi.org/10.1103/PhysRevLett.121.112001}{Phys.\ Rev.\ Lett.\  {\bf 121}, no. 11, 112001 (2018)}
  [\href{https://arxiv.org/abs/1803.02685}{\tt arXiv:1803.02685 [hep-lat]}].
  
  \bibitem{Lin:2018qky} 
  H.~W.~Lin, J.~W.~Chen, L.~Jin, Y.~S.~Liu, Y.~B.~Yang, J.~H.~Zhang and Y.~Zhao,
  ``Spin-Dependent Parton Distribution Function with Large Momentum at Physical Pion Mass,''
  \href{https://doi.org/10.1103/PhysRevLett.121.242003}{Phys.\ Rev.\ Lett.\  {\bf 121}, no. 24, 242003 (2018)}
  [\href{https://arxiv.org/abs/1807.07431}{\tt arXiv:1807.07431 [hep-lat]}].
  
\bibitem{Orginos:2017kos} 
  K.~Orginos, A.~Radyushkin, J.~Karpie and S.~Zafeiropoulos,
  Lattice QCD exploration of parton pseudo-distribution functions,
  \href{https://doi.org/10.1103/PhysRevD.96.094503}{Phys.\ Rev.\ D {\bf 96},  094503 (2017)}
  [\href{https://arxiv.org/abs/1706.05373}{\tt arXiv:1706.05373 [hep-ph]}].
  
  \bibitem{Lin:2017snn} 
  H.~W.~Lin {\it et al.},
  Parton distributions and lattice QCD calculations: a community white paper,
  \href{https://arxiv.org/abs/1711.07916}{\tt arXiv:1711.07916 [hep-ph]}.
  
  
\bibitem{Monahan:2018euv} 
  C.~Monahan,
  ``Recent Developments in $x$-dependent Structure Calculations,''
  PoS LATTICE {\bf 2018}, 018 (2018)
  [\href{https://arxiv.org/abs/1811.00678}{\tt arXiv:1811.00678 [hep-lat]}].

\bibitem{Cichy:2018mum} 
  K.~Cichy and M.~Constantinou,
  ``A guide to light-cone PDFs from Lattice QCD: an overview of approaches, techniques and results,''
  \href{https://arxiv.org/abs/1811.07248}{\tt arXiv:1811.07248 [hep-lat]}.
  
  
  \bibitem{Best:1997qp} 
  C.~Best {\it et al.},
  ``Pion and rho structure functions from lattice QCD,''
  \href{https://doi.org/10.1103/PhysRevD.56.2743}{Phys.\ Rev.\ D {\bf 56}, 2743 (1997)}
  [\href{https://arxiv.org/abs/hep-lat/9703014}{\tt arXiv:hep-lat/9703014}].
  
  
  \bibitem{Detmold:2003tm} 
  W.~Detmold, W.~Melnitchouk and A.~W.~Thomas,
  ``Parton distribution functions in the pion from lattice QCD,''
  \href{https://doi.org/10.1103/PhysRevD.68.034025}{Phys.\ Rev.\ D {\bf 68}, 034025 (2003)}
  [\href{https://arxiv.org/abs/hep-lat/0303015}{\tt arXiv:hep-lat/0303015}].
  
  
  \bibitem{Guagnelli:2004ga} 
  M.~Guagnelli {\it et al.} [Zeuthen-Rome (ZeRo) Collaboration],
  ``Non-perturbative pion matrix element of a twist-2 operator from the lattice,''
  \href{https://doi.org/10.1140/epjc/s2005-02121-5}{Eur.\ Phys.\ J.\ C {\bf 40}, 69 (2005)}
  [\href{https://arxiv.org/abs/hep-lat/0405027}{\tt arXiv:hep-lat/0405027}].
  
  
  \bibitem{Capitani:2005jp} 
  S.~Capitani, K.~Jansen, M.~Papinutto, A.~Shindler, C.~Urbach and I.~Wetzorke,
  ``Parton distribution functions with twisted mass fermions,''
  \href{https://doi.org/10.1016/j.physletb.2006.02.047}{Phys.\ Lett.\ B {\bf 639}, 520 (2006)}
  [\href{https://arxiv.org/abs/hep-lat/0511013}{\tt arXiv:hep-lat/0511013}].
  
  
  \bibitem{Bali:2013gya} 
  G.~Bali, S.~Collins, B.~Glässle, M.~Göckeler, N.~Javadi-Motaghi, J.~Najjar, W.~Söldner and A.~Sternbeck,
  ``Pion structure from lattice QCD,''
  \href{https://doi.org/10.22323/1.187.0447}{PoS LATTICE {\bf 2013}, 447 (2014)}
  [\href{https://arxiv.org/abs/1311.7639}{\tt arXiv:1311.7639 [hep-lat]}].
  
  
  \bibitem{Abdel-Rehim:2015owa} 
  A.~Abdel-Rehim {\it et al.},
  ``Nucleon and pion structure with lattice QCD simulations at physical value of the pion mass,''
  \href{https://doi.org/10.1103/PhysRevD.93.039904}{Phys.\ Rev.\ D {\bf 92}, no. 11, 114513 (2015)}
  \href{https://doi.org/10.1103/PhysRevD.93.039904}{Erratum: [Phys.\ Rev.\ D {\bf 93}, no. 3, 039904 (2016)]}
  [\href{https://arxiv.org/abs/1507.04936}{\tt arXiv:1507.04936 [hep-lat]}].
  
  

  
  
  \bibitem{Oehm:2018jvm} 
  M.~Oehm {\it et al.},
  ``$\langle x\rangle$ and $\langle x^2\rangle$ of the pion PDF from Lattice QCD with $N_f=2+1+1$ dynamical quark flavours,''
  \href{https://doi.org/10.1103/PhysRevD.99.014508}{Phys.\ Rev.\ D {\bf 99}, no. 1, 014508 (2019)}
  [\href{https://arxiv.org/abs/1810.09743}{\tt arXiv:1810.09743 [hep-lat]}].
  
  
\bibitem{Karpie:2018zaz} 
  J.~Karpie, K.~Orginos and S.~Zafeiropoulos,
  ``Moments of Ioffe time parton distribution functions from non-local matrix elements,''
  \href{https://doi.org/10.1007/JHEP11(2018)178}{JHEP {\bf 1811}, 178 (2018)}
  [\href{https://arxiv.org/abs/1807.10933}{\tt arXiv:1807.10933 [hep-lat]}].
  
  
  
  
 \bibitem{Briceno:2017cpo} 
  R. A. Brice\~no, M.~T.~Hansen and C.~J.~Monahan,
  ``Role of the Euclidean signature in lattice calculations of quasidistributions and other nonlocal matrix elements,''
  \href{https://doi.org/10.1103/PhysRevD.96.014502}{Phys.\ Rev.\ D {\bf 96}, no. 1, 014502 (2017)}
  [\href{https://arxiv.org/abs/1701.07465}{arXiv:1703.06072 [hep-lat]}].
  
  
  \bibitem{Ioffe:1969kf} 
  B.~L.~Ioffe,
  ``Space-time picture of photon and neutrino scattering and electroproduction cross-section asymptotics,''
  \href{https://doi.org/10.1016/0370-2693(69)90415-8}{Phys.\ Lett.\  {\bf 30B}, 123 (1969)}.
  
  

  
  \bibitem{lattices}
R.~Edwards, B.~Jo\'o, K.~Orginos, D.~Richards, and F.~Winter
``U.S. 2+1 flavor clover lattice generation program,"
Unpublished (2016).

\bibitem{rhmc} 
  S.~Duane, A.~D.~Kennedy, B.~J.~Pendleton and D.~Roweth,
  ``Hybrid Monte Carlo,''
  \href{https://doi.org/10.1016/0370-2693(87)91197-X}{Phys.\ Lett.\ B {\bf 195}, 216 (1987)}.
  
  
  
  \bibitem{Allton:1993wc} 
  C.~R.~Allton {\it et al.} [UKQCD Collaboration],
  ``Gauge invariant smearing and matrix correlators using Wilson fermions at Beta = 6.2,''
  \href{https://doi.org/10.1103/PhysRevD.47.5128}{Phys.\ Rev.\ D {\bf 47}, 5128 (1993)}
  [\href{https://arxiv.org/abs/hep-lat/9303009}{\tt arXiv:hep-lat/9303009}].
  
\bibitem{Bali:2016lva} 
  G.~S.~Bali, B.~Lang, B.~U.~Musch and A.~Schäfer,
  ``Novel quark smearing for hadrons with high momenta in lattice QCD,''
  \href{https://doi.org/10.1103/PhysRevD.93.094515}{Phys.\ Rev.\ D {\bf 93}, no. 9, 094515 (2016)}.
  [\href{https://arxiv.org/abs/1602.05525}{\tt arXiv:1602.05525 [hep-lat]}].
  

  
\bibitem{Detmold:2005gg} 
  W.~Detmold and C.~J.~D.~Lin,
  ``Deep-inelastic scattering and the operator product expansion in lattice QCD,''
  \href{https://doi.org/10.1103/PhysRevD.73.014501}{Phys.\ Rev.\ D {\bf 73}, 014501 (2006)}.
  [\href{https://arxiv.org/abs/hep-lat/0507007}{\tt  arXiv:hep-lat/0507007}].
  
  
  \bibitem{Karpie:2019eiq} 
  J.~Karpie, K.~Orginos, A.~Rothkopf and S.~Zafeiropoulos,
  ``Reconstructing parton distribution functions from Ioffe time data: from Bayesian methods to Neural Networks,''
  \href{https://arxiv.org/abs/1901.05408}{\tt arXiv:1901.05408 [hep-lat]}.
  
 
  
  
  \bibitem{ROOT} 
  Rene Brun and Fons Rademakers,
``ROOT - An Object Oriented Data Analysis Framework,"
\href{https://root.cern.ch/publications}{Proceedings AIHENP'96 Workshop, Lausanne, Sep. 1996, Nucl. Inst. \& Meth. in Phys. Res. A 389 (1997) 81-86.}


  \bibitem{Briceno:2018lfj} 
  R. A. Brice\~no, J.~V.~Guerrero, M.~T.~Hansen and C.~J.~Monahan,
  ``Finite-volume effects due to spatially nonlocal operators,''
  \href{https://doi.org/10.1103/PhysRevD.98.014511}{Phys.\ Rev.\ D {\bf 98}, no. 1, 014511 (2018)}
  [\href{https://arxiv.org/abs/1805.01034}{\tt arXiv:1805.01034 [hep-lat]}].


\bibitem{Chen:2018fwa} 
  J.~W.~Chen, L.~Jin, H.~W.~Lin, Y.~S.~Liu, A.~Schäfer, Y.~B.~Yang, J.~H.~Zhang and Y.~Zhao,
  ``First direct lattice-QCD calculation of the $x$-dependence of the pion parton distribution function,''
 \href{https://arxiv.org/abs/1804.01483}{\tt  arXiv:1804.01483 [hep-lat]}.
 
 
\bibitem{Karthik:2018wmj} 
  N.~Karthik, T.~Izubichi, L.~Jin, C.~Kallidonis, S.~Mukherjee, P.~Petreczky, C.~Shugert and S.~Syritsyn,
  ``Renormalized quasi parton distribution function of pion,''
  PoS LATTICE {\bf 2018}, 109 (2018)
  [\href{https://arxiv.org/abs/1811.06075}{\tt arXiv:1811.06075 [hep-lat]}].







\end{thebibliography}
\end{document}